\documentstyle[12pt,twoside]{article}

\catcode`\@=11
\def\newsymbol#1#2#3#4#5{\let\next@\relax%
 \ifnum#2=\@ne\else%
 \ifnum#2=\tw@\let\next@\msyfam@\fi\fi%
 \mathchardef#1="#3\next@#4#5}%"
\def\mathhexbox@#1#2#3{\relax%
 \ifmmode\mathpalette{}{\m@th\mnnathchar"#1#2#3}%"
 \else\leavevmode\hbox{$\m@th\mathchar"#1#2#3$}\fi}%"
\font\tenmsy=msbm10
\font\sevenmsy=msbm7
\font\fivemsy=msbm5
\newfam\msyfam
\textfont\msyfam=\tenmsy
\scriptfont\msyfam=\sevenmsy
\scriptscriptfont\msyfam=\fivemsy
\edef\msyfam@{\hexnumber@\msyfam}
\def\Bbb#1{\fam\msyfam\relax#1}
\catcode`\@=\active
\load{\footnotesize}{\sf}

\newtheorem{theorem}{Theorem}[section]
\newtheorem{proposition}[theorem]{Proposition}
\newtheorem{lemma}[theorem]{Lemma}
\newtheorem{corollary}[theorem]{Corollary}

\newtheorem{remark}[theorem]{Remark}

\newcommand{\bl}[1]{\begin{lemma}\label{#1}}
\newcommand{\el}{\end{lemma}}
\newcommand{\bt}[1]{\begin{theorem}\label{#1}}
\newcommand{\et}{\end{theorem}}
\newcommand{\bp}[1]{\begin{proposition}\label{#1}}
\newcommand{\ep}{\end{proposition}}
\newcommand{\bc}[1]{\begin{corollary}\label{#1}}
\newcommand{\ec}{\end{corollary}}

\newcommand{\eq}[1]{\begin{equation}\label{#1}}
\newcommand{\en}{\end{equation}}
\newcommand{\eqn}{\begin{eqnarray*}}
\newcommand{\enn}{\end{eqnarray*}}
\newcommand{\eqnn}{\begin{eqnarray}}
\newcommand{\ennn}{\end{eqnarray}}
\newcommand{\bi}{\begin{description}}
\newcommand{\ei}{\end{description} }
\newcommand{\non}{\nonumber}

\newcommand{\proof}{{\noindent \it Proof:\ }}
\newcommand{\qed}{\hfill $\Box$\par\medskip}

\newcommand{\RRR}[3]{(\pp{#1}, \q{#2}\pp{#3})_\hhh}

\newcommand{\pp}[1]{\Phi^\mu_{(#1)}}
\newcommand{\q}[1]{H_{#1}}
\newcommand{\wh}[1]{\tilde{H}_{(#1)}}
\newcommand{\kak}[1]{(\ref{#1})}
\newcommand{\ppp}[1]{\phi_{(#1)}}
\newcommand{\C}[2]{\footnotesize {\lk\!\!\!\begin{array}{c} #1 \\ #2 \end{array}
\!\!\!\rk}}
\newcommand{\e}[1]{E_{(#1)}}
\newcommand{\es}{\epsilon} 
\newcommand{\Ee}{E(e,\es)}
\newcommand{\g}[1]{\varphi_{(#1)}}
\newcommand{\G}[1]{\Psi_{(#1)}^\mu}
\newcommand{\Gt}[1]{\Psi_{(#1)}^3}
\newcommand{\dg}[1]{d\Gamma(#1)}
\newcommand{\ikm}{\int\! d{\bf k} \frac{1}{|k_1| |k_2|}}
\newcommand{\ikmm}[1]{\int
     _{\tiny \begin{array}{c}
             \kappa/m\leq|k_1|, |k_2|\leq \La/m \\
              |k_1|/|k_2|\ll 1
              \end{array}
       }#1 dk_1dk_2}%
\newcommand{\KK}{{\cal K}}
\newcommand{\KKK}{\hat{{\cal K}}}
\newcommand{\ab}[1]{a_{#1}\varphi_-+b_{#1}\varphi_+}
\newcommand{\lk}{\left(}
\newcommand{\rk}{\right)}
\newcommand{\lkk}{\left\{}
\newcommand{\rkk}{\right\}}
\newcommand{\BR}{{{\Bbb R}^3}}
\newcommand{\RR}{{{\Bbb R}}}
\renewcommand{\l}[1]{[\log(\La/m)]^{#1}}
\newcommand{\pr}{\epsilon_0} 
\newcommand{\hi}{H_{\rm I}}
\newcommand{\hii}{H_{\rm II}}
\newcommand{\z}{H_0^{\!-1}}
\newcommand{\pfa}{\pf  A} 
\newcommand{\bs}{\lk\!-\frac{\s  B}{2}\!\rk} 
\newcommand{\sbb}{ \s  B^+} 
\newcommand{\cf}[1]{{\cal F}^{(#1)}}
\newcommand{\dz}{{\cal D}_a} 
\newcommand{\dgg}{{\cal D}_g}

\newcommand{\fg}{f_{\rm g}}
\newcommand{\BC}{{{\Bbb C}}}
\newcommand{\CC}{{{\Bbb C}}}
\newcommand{\La}{\Lambda} 
\newcommand{\liml}{\lim_{\Lambda\rightarrow\infty}}

\renewcommand{\d}{\displaystyle}
\newcommand{\LRJ}{{L^2(\BR\!\times\!\{1,2\})}}
\newcommand{\LR}{{L^2(\BR)}}
\newcommand{\fff}{{{\cal F}}}
\newcommand{\fkd}{\fff_\kappa}
\newcommand{\fk}{\fff_0}
\newcommand{\ffff}{{{\cal F}_{\rm fin}}}
\newcommand{\hhh}{{{\cal H}}}

\newcommand{\hk}{{\hhh_\kappa}}
\newcommand{\is}{\inf\sigma}
\newcommand{\f}{^{-1}}
\newcommand{\wick}[1]{:\!{#1}\!:}
\newcommand{\add}{a^{\ast}}
\newcommand{\ass}{a^\sharp}
\newcommand{\st}{{\rm s}\!-\!}
\newcommand{\hf}{H_{\rm f}}
\newcommand{\pf}{{ P_{\rm f}}}
\newcommand{\mmm}{\sum_{\mu=1}^3}
\newcommand{\gr}{\varphi_{\rm g}}
\newcommand{\kp}{(e,\es)}
\newcommand{\he}{H(e,\epsilon)} 
\newcommand{\kr}{\varphi_{{\rm g},\kappa}}
\newcommand{\grr}{\gr}
\newcommand{\half}{\frac{1}{2}}
\newcommand{\halfm}{\frac{1}{2m}}
\newcommand{\han}{{1/2}}
\newcommand{\ehalfm}{\frac{e}{2m}}
\newcommand{\vp}{\hat{\varphi}}
\newcommand{\vpm}{{\hat{\varphi}_m}}
\newcommand{\av}{{A_{\vp}}}
\newcommand{\bv}{{B_{\vp}}}
\newcommand{\EE}{\d\lk\!-\frac{\e 2}{2}\!\rk}
\newcommand{\EEE}{ \e 2/2} 
\newcommand{\ce}{{\cal  E}}
\newcommand{\ccr}{{\cal R}}
\newcommand{\skk}{\s(k_1,k_2)}
\renewcommand{\AA}{E_{a}}
\newcommand{\AB}[1]{\langle #1 \rangle_E}
\newcommand{\BB}{E_{b}}
\newcommand{\mass}{m_{\rm eff}}
\newcommand{\ph}{m_{\rm ph}} 
\newcommand{\jj}{\sum_{j=1,2}}
\newcommand{\s}{\sigma}
\begin{document}

\makeatletter
\@addtoreset{equation}{section}
\makeatother
\def\theequation{\arabic{section}.\arabic{equation}}

\pagestyle{headings}

\setlength{\baselineskip}{16pt}
\title {Mass Renormalization in \\
 Non-relativistic Quantum Electrodynamics\\
      with Spin $\han$}
\author{F. Hiroshima\thanks{hiroshima@mpg.setsunan.ac.jp} 
    and K. R. Ito\thanks{ito@mpg.setsunan.ac.jp}}

%%%%%%%%%%%%%%%%%%%%%%%%%%%%%%%%%%%%%%%%
%% 2004 July       created by Hiro
%% 2004 Sept  04   handed to spohn
%% 2004 Nov   18   final ? edition by Hiro
%% 1207   2004 Dec   05-06 Appendix Added by Ito
%% 1207'  2004 Dec   08    Corrected by Hiro  
%% 1210   2004 Dec   09-10 Corrected by Ito
%%%%%%%%%%%%%%%%%%%%%%%%%%%%%%%%%%%%%%%%

\date{\today}

\maketitle

\begin{center}
Department of Mathematics and Physics,\\
Setsunan University \\ 
Neyagawa, Osaka,  572-8508, \\
Japan
\end{center}

\begin{abstract}
The effective mass $\mass$ of the the Pauli-Fierz Hamiltonain 
with ultraviolet cutoff $\La$ and the bare mass $m$ 
in nonrelativistic QED  with spin $\han$ is investigated. 
Analytic properties of $\mass$ in coupling constant $e$ are shown 
and explicit forms of constants $a_1(\La/m)$ and $a_2(\La/m)$ 
depending on  $\La/m$  such that 
$$\mass/m =1 + a_1(\La/m) e^2+ a_2(\La/m) e^4+ {\mathcal O}(e^6)$$
are  given. 
It is shown that the spin interaction enhances the effective mass and 
that there exist strictly positive constants 
$b_1,b_2, c_1$ and $c_2$ such that 
$$\d 
b_1\leq \liml \frac{a_1(\La/m)}{\log (\La/m)}\leq b_2,\ \ \ 
  -c_1\leq \liml \frac{a_2(\La/m)}{(\La/m)^2}\leq -c_2.$$ 
In particular  $a_2(\La/m)$ does not 
diverges as $\pm [\log(\La/m)]^2$ but $-(\La/m)^2$. 
\end{abstract}

\setlength{\baselineskip}{20pt}

\section{Introduction}
One electron with the bare mass $m$ interacting with a quantized
radiation field carries a virtual cloud of photons. 
This virtual cloud enhances the bare mass 
$m$ to an effective mass $\mass\geq m$.  
In this paper the effective mass is defined as 
the inverse of the Hessian  of 
the ground state energy of the Hamiltonian at total momentum zero. 
The electron includes spin and the quantized radiation field has 
an infrared cutoff $\kappa$ and an ultraviolet cutoff $\La$. 
We are interested in the asymptotic behavior of $\mass$ 
as ultraviolet cutoff $\La$ goes to infinity. 
It is admittedly a subtle problem.  Our contribution in this 
paper is to investigate the asymptotics of the fourth order of 
a coupling constant $e$ of the effective mass $\mass$. 
Although the purpose of this paper is to prove the same things as 
the spinless case established in \cite{hisp2}, 
the result is completely different from it. 
In the spinless case its  divergence is $\sqrt\La$ and 
in the present case $\La^2$ as $\La\rightarrow \infty$. 

It is proven in \cite{arai1} that the authentic Pauli-Fierz model 
in non-relativistic QED can be 
derived from QED by a  scaling limit. 
QED is renormalizable, 
and then the Pauli-Fierz model may be expected to be 
also renormalizable. In facts, there exists  no fermion loop but only 
photon absorption-emission diagrams  in the non-relativistic QED. 
Although QED is believed to be trivial after the renormalization, 
there is a strong  belief that the Pauli-Fierz model with spin
is renormalizable and the renormalized theory is non-trivial. 
Nevertheless we  show that 
there exist divergences in higher order perturbation expansions, 
which cannot be removed by the conventional renormalization scheme.
So it is inconceivable that 
the effective mass $\mass$ of the Pauli-Fierz Hamiltonian 
with spin can be renormalized in our procedure. 

In the spinless case, the  $e^2$ order of the ground state energy 
$\d E=\sum_{n=0}^\infty \frac{e^n}{n!} \e n$ of 
the Pauli Fierz Hamiltonian with total momentum zero 
is identically zero, the inclusion of spin, however,  presents 
that $\d \half E_{(2)}\not=0$ and it diverges like
$\Lambda^{2}$ as $\La$ goes to infinity. 
This divergence  can 
be hidden as a constant to define a Wick  product. 
It begins  to appear, however,  in the connected diagrams 
of higher orders. 
Since the Pauli-Fierz Hamiltonian  before the scaling limit 
must have substantially many cutoffs  which 
may be shielded by the scaling limit, 
the appearance of such divergences in higher order diagrams
is not a contradiction. In fact without these cutoffs, 
the scaling limit would not exist. 
As a consequence $E_{(2)}$ causes $e^4$ order of $\mass$ to 
include $\La^2$ divergence. 

The effective mass  and the ground state energy of 
the Pauli-Fierz Hamiltonian are  studied in 
e.g., \cite{caha, ch, cvv,   hasi, hisp2, hvv, li, lilo2, sp5}. 
We note that a precise grasp of the definition of the effective 
mass $\mass$ is also a problem.  Actually there are several 
ways to define the effective mass. 
Alternative definition of the effective mass is given in 
e.g., Lieb-Loss \cite{lilo2} and Lieb \cite{li}. It is worth checking
that both of definitions coincide with each others under some 
conditions, e.g., $\La\ll 1$.  Catto-Hainzl \cite{ch} expand the 
ground state energy  of the Pauli-Fierz Hamiltonian with spin and
without infrared cutoff up to $e^4$ order. Hainzl and Seiringer 
\cite{hasi} give an exact form of the coefficient of $e^2$ 
order of the effective mass. 
Moreover other related work is  Spohn \cite{sp5} where upper and 
lower bounds of the effective mass of the Nelson model are studied. 
See also \cite{sp6} for recent movement and progress in this field.

Our results are summarized as follows:\\
{\bf Theorem A} (Analytic properties) {The effective mass $\mass$ divided by the bare mass $m$, 
$\mass/m$,  is a function of $e^2$ and $\La/m$,  and   analytic in $e^2$
for  $e^2<e_{00}$  with some $e_{00}>0$ 
depending on $\La$.

\noindent 
{\bf Theorem B} ($(\La/m)^2$-divergence){
Let 
$$ \frac{\mass}{m}=1+a_{1}(\La/m)e^{2}
                     +a_{2}(\La/m)e^{4} +{\mathcal  O}(e^{6}).$$ }
Then there  exist strictly positive constants $b_{i}$ and 
$c_{i}$, $i=1,2$,  such that
$$
b_1\leq \liml \frac{a_1(\La/m)}{\log (\La/m)}\leq b_2, \quad 
-c_1\leq \liml \frac{a_2(\La/m)}{(\La/m)^2}\leq -c_2.$$

\noindent
{\bf Theorem C} (Enhancement by spin) {Let $\mass^{\rm spinless}$ be the effective mass for 
the spinless Pauli-Fierz Hamiltonian. Then   
$\mass>\mass^{\rm spinless}$ 
follows for $e$ with a sufficiently small $|e|$ but $\not=0$. 
}

Theorem A,  B and C are 
converted to Corollary \ref{an}, Theorem \ref{main} and Corollary \ref{p1}, 
respectively. 
 We organize the paper as follows: Section 2 is devoted to 
expanding $\mass$ up to $e^4$. 
In section 3 the divergence of the forth order of $\mass$  is estimated. 
Finally in section 4, some additional arguments on the scaling limit 
and open problems on this model are given.

%%%%%%%%%%%%%%%%%%%%%%%%%%%%%%%%%%%%%%%%%%%%%%%%%%%%%%
\subsection{Definitions of the non-relativistic QED}
%%%%%%%%%%%%%%%%%%%%%%%%%%%%%%%%%%%%%%%%%%%%%%%%%%%%%
Let $\fff$ be the boson Fock space over $\LRJ$ given by 
$$\displaystyle 
    \fff\equiv \bigoplus_{n=0}^\infty\left [\otimes_s^n\LRJ\right],$$ 
where 
$\otimes_s^n$ denotes the $n$-fold symmetric tensor product and 
$\otimes _s^0\LRJ\equiv \BC$. 
The Fock vacuum $\Omega\in\fff$ is defined by 
$\Omega\equiv\{1,0,0,..\}$. 
 Let  $a(f)$, $f\in\LRJ$, be  
the creation operator and 
 $\add(f)$, $f\in\LRJ$, the annihilation operator on $\fff$ defined by 
\eqn 
&& (\add(f)\Psi)^{(n+1)}\equiv \sqrt{n+1}S_{n+1}(f\otimes \Psi^{(n)}),\\
&& D(\add(f))=\lkk\Psi\in \fff| \sum_{n=0}^\infty(n+1)
     \|S_{n+1}(f\otimes \Psi^{(n)})\|_{\otimes^n\LRJ}^2<\infty\rkk, 
\enn
and $a(f)=[\add(\bar f)]^\ast$, where $S_n$ denotes the symmetrizer, 
 $D(X)$  the domain of operator $X$, and $\|\cdot\|_{\cal K}$ the norm 
on ${\cal K}$. The scalar product on ${\cal K}$ is denoted by 
$(f,g)_{\cal K}$ which is linear in $g$  and anti-linear in $f$.  
They satisfy canonical commutation relations: 
   $$[a(f), \add(g)]=(\bar f, g)_\LRJ,\ \ \ 
       [a(f), a(g)]=0, \ \ \  [\add(f), \add(g)]=0,$$
where 
$[X, Y]=XY-YX$. 
Formal kernels of $a(f)$ and $\add(f)$ are denoted by 
$a(k,j)$ and $\add(k,j)$, $(k,j)\in\BR\times\{1,2\}$, respectively.  
Then we write as $\jj \int \ass(k,j) f(k,j) dk$ for $\ass(f)$. 
The linear hull of 
$\{\add(f_1)\cdots \add(f_n)\Omega|f_j\in\LRJ,j\geq 0\}$ 
is dense in $\fff$. 
Let $T:\LR\rightarrow\LR$ be a self-adjoint operator. 
We define $\Gamma(e^{itT})$ by 
$$\Gamma(e^{itT}) \add(f_1)\cdots \add(f_n)\Omega\equiv 
\add(e^{itT}f_1)\cdots \add(e^{itT}f_n)\Omega.$$
Thus $\Gamma(e^{itT})$ turns out to be a strongly continuous 
one-parameter unitary group in $t$, 
which implies that there exists a self-adjoint operator $\dg T$ on 
$\fff$ such that 
$$\Gamma(e^{itT})=e^{it\dg T},\ \ \ t\in\RR.$$
The self-adjoint operator $\dg T$ is referred to the second 
quantization of $T$. We define a Hilbert space $\hhh$ by 
$$\hhh\equiv \CC^2\otimes\fff.$$
We identify $\hhh$ as 
$$\hhh\cong 
\bigoplus_{n=0}^\infty 
\left [
\CC^2\otimes(\otimes_s^n\LRJ)\right]\equiv \bigoplus_{n=0}^\infty \cf n.$$
The Pauli-Fierz Hamiltonian with total momentum 
$p=(p_1, p_2,p_3)\in\BR$  is given by a symmetric operator on $\hhh$: 
\eqn 
H_m(e,p)\equiv  \halfm  \lkk\sum_{\mu=1}^3 \sigma_\mu \otimes  (p_\mu-\pf_\mu-e\av_\mu)\rkk ^2+1\otimes \hf,\\
\enn 
where $m>0$ and $e\in\RR$ denote the mass and the charge of an electron, 
respectively, 
$\sigma\equiv (\sigma_1,\sigma_2,\sigma_3)$  
the $2\times2$ Pauli-matrices given by 
$$
\s_1\equiv \lk 
\begin{array}{cc}
0&1\\  
1&0
\end{array}\rk, \ \ \ 
\s_2\equiv \lk 
\begin{array}{cc}
0&-i\\ 
i&0
\end{array}\rk,\ \ \ 
\s_3\equiv \lk 
\begin{array}{cc}
1&0\\
0&-1\end{array}\rk,  
$$
and the free Hamiltonian $\hf$, the momentum operator $\pf$ and 
the quantum radiation field $\av_\mu$ are given by 
\eqn
&&\hf\equiv \dg{\omega},\ \ \ \pf_\mu\equiv \dg {k_\mu},\\
&&
\av_\mu\equiv 
\frac{1}{\sqrt2}\jj \int \frac{\vp(k)}{\sqrt{\omega(k)}}
     e_\mu(k,j) (\add(k,j)+a(k,j))dk, \ \ \ 
\mu=1,2,3.
\enn
Here 
$e(k,j)$, $j=1,2$, denotes polarization vectors such that 
three vectors $e(k,1)$, $e(k,2)$,  $k/|k|$ form 
a right-handed system in $\BR$, i.e., 
$$|e(k,j)|=1, \ \ \ j=1,2, \ \ \ e(k,1)\cdot e(k,2)=0,\ \ \ 
e(k,1)\times e(k,2)=k/|k|.$$ 
We fix polarization vectors, e.g., 
for $k=(|k| \cos\theta\cos\phi, |k|\sin\theta\cos\phi, |k| \cos\theta)$, 
$$e(k,1)= (-\sin\theta,\cos\theta,0),\ \ \
e(k,2)=(-\cos\theta\sin\phi,-\sin\theta\sin\phi,\cos\phi),
$$  where   
$0\leq \theta\leq2\pi$, $-\pi/2\leq \phi\leq \pi/2$.  
Note that $e(-k,1)=e(k,1)$, $e(-k,2)=-e(k,2)$. 
Since 
\eqn 
&& [\pf_\mu, \av_\nu]=\frac{1}{\sqrt2}\jj 
\int \frac{\vp(k)}{\sqrt{\omega(k)}}k_\mu e_\nu(k,j) (\add(k,j)-a(k,j))dk, 
\enn 
it follows that 
$$H_m(e,p)=
\halfm \lkk 
1\otimes \sum_{\mu=1}^3 (p_\mu-\pf_\mu-e\av_\mu)^2\rkk +1\otimes \hf -
\ehalfm \sum_{\mu=1}^3 \s_\mu  \otimes B_\mu,$$
where $B_\mu$ denotes the  quantum magnetic field given by 
$$
\bv_\mu\equiv 
\frac{i}{\sqrt2}\jj \int 
\frac{\vp(k)}{\sqrt{\omega(k)}}(k\times e(k,j))_\mu (\add(k,j)-a(k,j))dk. 
$$
We simply write $H_m(e,p)$ as 
$$H_m(e,p)=\halfm (p-\pf-e\av)^2+ \hf -\ehalfm \s  \bv,\ \ \ p\in\BR.$$
The spinless Pauli-Fierz Hamiltonian is given by 
$$H^{\rm spinless}_m(e,p)=\halfm (p-\pf-e\av)^2+ \hf$$
which acts on $\fff$. 
We assume that 
$\vp(k)=\overline{\vp(k)}=\vp(-k)$  
and 
$$\int_\BR(\omega(k)^{-2} +\omega(k))|\vp(k)|^2 dk<\infty.$$
Then $H_m(e,p)$ is self-adjoint on $D(\frac{1}{2m}\pf^2+\hf)$ for $(e,p)\in\RR\times\BR$ 
such that  $|e|<{\bf e}$ and $|p|<{\bf p}$ 
with some constants  ${\bf e}$ and ${\bf p}$. 
This is proven 
by the Kato-Rellich theorem and  
using the inequalities 
\eqn
&&\|\add(f)\Psi\|_\fff\leq \|f/\sqrt\omega\|_\LRJ 
\|\hf^\han\Psi\|_\fff +\|f\|_\LRJ\|\Psi\|_\fff,\\
&&\|a(f)\Psi\|_\fff\leq \|f/\sqrt\omega\|_\LRJ \|\hf^\han\Psi\|_\fff.
\enn
%%%%%%%%%%%%%%%%%%%%%%%%%%%%%%%%%%%%%%%%%%%%%%%
\subsection{Mass renormalization}
%%%%%%%%%%%%%%%%%%%%%%%%%%%%%%%%%%%%%%%%%%%%%%%
In what follows we set 
\eq{fix2}
 \vp(k)\equiv \lkk\begin{array}{ll}
0,& |k|<\kappa,\\
1/\sqrt{(2\pi)^3},& \kappa\leq |k|\leq \La,\\
0,&|k|>\La.
\end{array}
\right.
\en
The effective mass $\mass=\mass(e^2,\La,\kappa,m)$ is defined by 
$$
\frac{1}{\mass}\equiv \frac{1}{3}\Delta_p \is (H_m(e,p)) \lceil_{p=0}.$$
Removal of the ultraviolet cutoff $\Lambda$ through mass renormalization
means to find sequences $\{m\}$ and $\{\La\}$ such that 
$\Lambda\rightarrow \infty$ and $m\rightarrow 0$, 
and $\mass$ converges to  some positive constant. To achieve 
this, we want to find constants $\beta<0$ and $b>0$ such that 
\eq{m3}
\liml \mass(e^2,  \Lambda, \kappa \Lambda^\beta, (b\Lambda)^\beta)=\ph,
\en 
where $\ph>0$ is a given constant. We will see later that 
$\mass/m$ is a function of $e^2$, $\Lambda/m$ and $\kappa/m$. 
Set 
\eq{14}
f(e^2, \Lambda/m, \kappa/m)\equiv \frac{\mass}{m}. 
\en 
The analysis of \kak{m3}  can be reduced to find constants 
$0\leq \gamma<1$ and $0<b_0<\infty$ such that 
\eq{m2}
\liml  \frac{f(e^2, \Lambda/m, \kappa/m)}{(\Lambda/m)^\gamma} 
=b_0.
\en
If we succeed in finding constants $\gamma$ and $b_0$ such as 
in \kak{m2},  then, taking 
$$\beta\equiv \frac{-\gamma}{1-\gamma}<0,\ \ \ b\equiv 1/b_1^{1/\gamma},$$
we see that  
$$\liml\mass(e^2,  \Lambda, \kappa\Lambda^\beta, (b\Lambda)^\beta)
=
\liml 
b_0\lk \frac{\La}{b_1^{1/\gamma}}\rk^\beta \lk 
  \frac{\La}{(\La/b_1^{1/\gamma})^\beta}\rk^\gamma
     =    b_0b_1,  $$
where $b_1$ is a parameter, which is adjusted such as $b_0b_1=\ph$. 
Hence we will be able to establish \kak{m3}.
The  mass renormalization is, however,  a subtle problem, 
and unfortunately we can not yet find constants $\gamma$ and $b_0$ 
such as in \kak{m2}.  It is proven that $f$ is analytic in $e$ with 
$|e|$ sufficiently small under some conditions. 
Let us $\kappa>0$ be fixed and set 
$$f(e^2,\La/m,\kappa/m)=\sum_{n=0}^\infty a_n(\La/m) e^{2n}.$$
In the previous paper \cite{hisp2} it is proven  that 
\eq{yu}
a_2(\La/m)\approx \sqrt{\La/m},\ \ \ 
\La\rightarrow \infty,
\en 
for a spinless Pauli-Fierz Hamiltonian, where $X\approx Y$ means that 
there exist positive constants $x_1$ and $x_2$ such that 
$$x_1\leq \lim_{\La\rightarrow\infty}(X/Y)\leq x_2.$$ 
\kak{yu} suggests that $\gamma\geq 1/2$ for $e\not=0$, 
and that $\gamma$ is not analytic in $e$ since $\gamma=0$ for $e=0$.  
We are interested in knowing the dependence of spin for the asymptotics
of $a_2(\La/m)$ as $\La\rightarrow \infty$.   
Although it is desirable that $\gamma<1$, actually in this paper 
we shall show that 
$$a_2(\La/m)\approx  -(\La/m)^2,\ \ \ 
\La\rightarrow \infty.$$ 
This suggests that 
$\gamma\geq 2 $ and,  then mass renormalization such as 
\kak{m3} may not work. 

%%%%%%%%%%%%%%%%%%%%%%%%%%%%%%%%%%%%%%%%%%%%%%%%%%%%%
\section{Effective Mass and Analyticity}
%%%%%%%%%%%%%%%%%%%%%%%%%%%%%%%%%%%%%%%%%%%%%%%%%%%%%
\subsection{Analytic properties} 
Let $$T_m(e,p)\equiv \half(p-\pf-eA_{\vp_m})^2+\hf-\frac{e}{2}\s B_{\vp_m},$$
where 
\eq{fix}
\vp_m(k)\equiv \vp(mk)=\lkk
\begin{array}{ll}
   0,& |k|<\kappa/m,\\
   1/\sqrt{(2\pi)^3},& \kappa/m\leq |k|\leq \La/m,\\
   0,&|k|>\La/m.
\end{array}
\right.
\en
Since $\av\cong mA_\vpm$, $\bv\cong mB_\vpm$, 
$\hf\cong m\hf$ and $\pf\cong m\pf$, where $X\cong Y$ denotes 
the unitary equivalence between  $X$ and $Y$,  we see  that 
\eq{01}H_m(e,p)\cong mT_m(e,p/m).
\en 
The set of operators $J=(J_1,J_2,J_3)$ on $\hhh$ is  defined by
$$J_\mu \equiv \dg{(k\times(-i\nabla_k))_\mu}+\half\s_\mu,\ \ \ \mu=1,2,3.$$
In \cite{hisp3} it is written informally as 
\eqn
&& \dg{(k\times(-i\nabla_k))_\mu}
= -i\jj\int \add(k,j) (k\times \nabla_k) _\mu a(k,j) dk\\
&&\hspace{3cm} 
+i\int \frac{k_\mu}{|k|} \{\add(k,2)a(k,1)-\add(k,1)a(k,2)\}dk.
\enn 
Let $n\in\BR$ with $|n|=1$, $\theta\in\RR$, 
and ${\rm R}=(R_{\mu\nu})_{1\leq\mu,\nu\leq 3}$ be the $3\times 3$
matrix describing the rotation around $n$ through  angle $\theta$. 
Since 
$ 
e({\rm R}k,j)={\rm R}e(k,j)$ for $ k\not= |k| n$, $j=1,2$, 
we have 
\eqn 
&& e^{i\theta n\cdot J}A_\mu e^{-i\theta n\cdot J}=
\sum_{\nu=1}^3R_{\mu\nu}A_\nu, \ \ 
e^{i\theta n\cdot J}\pf_\mu e^{-i\theta n\cdot J}=
\sum_{\nu=1}^3R_{\mu\nu}\pf_\nu,\ \ 
e^{i\theta n\cdot J}\hf e^{-i\theta n\cdot J}=\hf,\\
&&e^{i\theta n\cdot J} B_\mu e^{-i\theta n\cdot J}=
\sum_{\nu=1}^3R_{\mu\nu}B_\nu, \ \ 
e^{i\theta n\cdot J} \s_\mu e^{-i\theta n\cdot J}=
\sum_{\nu=1}^3R_{\mu\nu}\s_\nu.
\enn 
Hence  
$$e^{i\theta n\cdot J} T_m(e,p) e^{-i\theta n\cdot J}=T_m(e,{\rm R}\f p).$$
In particular for $p\not=(0,0,p_3)$, 
\eqnn 
\label {g1}
e^{i\theta_p n_p \cdot J}T_m(e,p)  e^{-i\theta_p n_p\cdot J}=T_m(e,|p|n_z),
\ennn
where 
$\theta_p=\arccos\lkk (n_z,p)/|p|\rkk$, $n_z=(0,0,1)$, and $n_p=p\times n_z/|p\times n_z|$. 
It is also seen that for $p\in\BR$, 
$$T_m(e,|p|n_z)\cong T_m(e,-|p|n_z).$$
Then 
$$H_m(e,p)\cong m{T_m(e,(|p|/m)n_z)},\ \ \ p\in\BR.$$
Let 
$\wick{X}$ be the Wick product of $X$. 
We define 
$$\he \equiv \wick{T_m(e,\epsilon n_z)},\ \ \ \epsilon\in\RR.$$
Set 
$$E(e,\es)\equiv \is(\he ).$$ 
Since 
$$ 
T_m(e,\epsilon n_z)=\he +e^2 \frac{2}{3}\|\vp/\sqrt{2\omega}\|_\LR^2
$$ 
and 
$$\is(H_m(e,p))=\is(mT_m(e,(|p|/m)n_z))
    = m E(e,|p|/m)-e^2\frac{2}{3}m\|\vp/\sqrt{2\omega}\|^2_\LR,$$
we have 
$$\frac{m}{\mass}=\partial_\es^2 E(e,\es)\lceil_{\es=0}.$$
Then  it is  also seen that $m/\mass$ depends on $\La/m$ and $\kappa/m$. 
We study $\wick{\he }$ for $H_m(e,p)$ for simplicity.  
We set $A\equiv A_\vpm$ and $B\equiv B_\vpm$. 

We review analytic properties  of both of $\Ee$ and a ground state
of $\he$. In \cite{hisp2} the two fold degeneracy of a ground state
has been proven for the case where $\omega(k)=\sqrt{|k|^2+\nu^2}$, 
$\nu>0$, and $\kappa=0$. With a small modification of \cite{hisp2} 
we can also prove the two fold degeneracy of a ground state 
for our case, i.e., $\omega(k)=|k|$ and $\kappa>0$, and show 
its analytic properties. As is seen below in our case $E\kp$ i
s not an isolated eigenvalue and is degenerate. 
Then it is {\it not} clear that a ground state of $\he$ and 
$E\kp$ are analytic in $e$ and $\es$. Let 
$$\varphi_+\equiv  \C 1 0\otimes \Omega,\ \ \ \ 
\varphi_-\equiv  \C 0 1 \otimes \Omega.
$$ 
We shall prove the following lemma.  
\bl{anal}
There exist constants $e_0>0$ and $\pr>0$ such that 
for 
$(e,\es)\in 
\dz\equiv \{(e,\es)\in\RR^2||e|<e_0, |\es|<\pr\}$, 
(1) the dimension of ${\rm Ker} (\he-\Ee)$ is two, 
(2) $\Ee$ is an analytic function of $e^2$ and $\es^2$, 
(3) there exists a  strongly analytic ground state of $\he$.
\el
From this lemma the following corollary immediately follows. 
\bc{an}
Effective mass $\mass$ is analytic of $e^2$ 
on $\{e\in\BR|e^2<e_{00}\}$ with some $e_{00}>0$.
\ec
To prove Lemma \ref{anal}  we need several lemmas. 
Let 
$$\fk\equiv \fff(L^2(\RR^3_\kappa\times\{1,2\})),\ \ \ 
\fkd \equiv \fff(L^2({\RR^3_\kappa}^\perp\times\{1,2\})),
$$
where $\RR^3_\kappa=\{k\in\BR||k|\leq \kappa\}$. 
Then it follows that 
$
\fff\cong \fkd \otimes \fk
$
and 
\eq{k}
\hhh\cong \hk\otimes \fk,
\en 
where $\hk\equiv \CC^2\otimes\fkd $.
It is seen that 
${\he}$ is reduced by $\hk$. 
Let 
$$K(e,\es)\equiv {\he}\lceil_{\hk}.$$
Under the identification \kak{k},  we have 
\eq{re}
{\he}\cong K\kp\otimes 1_{\fk}
+1_{\hk}\otimes (\hf\lceil_{\fk}).
\en 
Since 
$\s(\hf\lceil_{\fk})=[0,\infty)$, 
we have by \kak{re},  
$$\is(K(e,\es))=\Ee.$$
\bp{f1}
(1) It follows that 
$$\is_{\rm ess} (K(e, |p|) )-\is(K(e, |p|))
\geq 
\inf_{|k|>\kappa} \lkk E(e, |p-k|)+\omega(k)-E(e, |p|)\rkk.$$ 
In particular, assume that  
$\inf_{|k|>\kappa} \lkk E(e, |p-k|)+\omega(k)-E(e, |p|)\rkk>0$. 
Then 
$K(e, \pm|p|)$ has a ground state in $\hk$. 
(2) The dimension of ${\rm Ker}\ (K(e, \pm|p|)-\is(K(e, \pm|p|))$ is two. 
\ep
\proof 
(1) and (2) follow in the similar  manner to \cite[Theorem 2.3]{fr2} and 
\cite{hisp3}, respectively. 
\qed
Since $\Ee$ is continuous in $(e, \es)$,  and 
$\d E(e, |p-k|)+\omega(k)-E(e, |p|)=|k|$ for $(e, p)=(0,0)\in\RR\times\BR$, 
it is seen that 
$$\lkk (e, p)\in\RR\times\BR \left| \inf_{|k|>\kappa} 
\lkk E(e,  |p-k|)+\omega(k)-E(e, |p|)\rkk >0\right. \rkk \not=
\emptyset.$$ 
Thus by Proposition \ref{f1} we have the corollary. 
\bc{f2}
There exist constants $\es^\ast>0$ and $e^\ast>0$ such that 
$K(e, \es)$ has a two-fold ground state for
$(e,\es)\in\dgg\equiv \{(e,\es)\in\RR^2||e|<e^\ast, |\es|<\es^\ast\}$. 
\ec
Let $\kr\kp\in\hk$ be a ground state of $K\kp$, i.e., 
$K\kp\kr\kp=E\kp\kr\kp$. 
Since 
$$ e^{i\theta n_z\cdot J} \he e^{-i\theta n_z\cdot J}
        =\he,\ \ \ \theta\in\BR,$$
and $e^{i\theta n_z\cdot J}$ is reduced  by  $\hk$, 
it follows that 
\eqnn 
\label {g2} 
e^{i\theta (p/|p|)\cdot J}K\kp e^{-i\theta (p/|p|) \cdot J}=K\kp
\ennn 
on $\hk$. 
Note that 
$\sigma(n_z\cdot J)={\Bbb Z}_\han\equiv \{\pm 1/2, \pm 3/2,....\}$ and
$\sigma(n_z\cdot J\lceil_{\hk})\subset {\Bbb Z}_\han$. 
Then  $K\kp$ and $\hk$ are decomposed as 
\eqn 
\hk=\bigoplus_{z\in \sigma(n\cdot J\lceil_{\hk})} \hk(z), \ \ \ \ 
K\kp=\bigoplus_{z\in \sigma(n\cdot J\lceil_{\hk})} K_z\kp.
\enn 
Note that 
$\hk(z)$ is independent of $\es$. 
Let 
$$\displaystyle 
K_-\kp\equiv \!\!\!\bigoplus_{z\leq -\han} K_z\kp,\ \ \ 
\displaystyle  K_+\kp\equiv \!\!\!\bigoplus_{z\geq \han}K_z\kp.
$$ 
Then 
$$K\kp=K_-\kp\oplus K_+\kp.$$ 
\bl{tani}
Assume that $(e,\es)\in \dgg$. 
Then 
$K_\pm\kp$ has a unique ground state ${\kr}_\pm\kp$ such that 
${\kr}_\pm\kp\in \hk(\pm\han)$,  and 
$\is(K_\pm\kp)=E\kp$ follows, i.e., 
\eqn&&
{\rm Ker}(K\kp-E\kp)\\
&&=
[\hk(-\han)\cap {\rm Ker}(K\kp-E\kp)] 
     \oplus[\hk(\han)\cap {\rm Ker}(K\kp-E\kp)].
\enn 
\el 
\proof 
It is proven in the similar manner as \cite[Theorem 1.2]{hisp3}. 
\qed
\bl{hio}
For $(e,\es)\in\dgg$, 
$\is(K_\pm\kp)$ is an isolated eigenvalue.
\el
\proof 
By Proposition \ref{f1}, 
$\delta\equiv \is_{\rm ess}(K\kp)-\is(K\kp)>0$, which implies that 
$\is_{\rm ess}(K_\pm\kp)-\is(K_\pm\kp)\geq \delta>0$. Then the lemma follows. 
\qed

Now we are in the position to show analytic properties of $\kr$ on  $e$. 
\bl{ana}
There exist constants $e_0>0$  and $\pr>0$ such that 
${\kr}_\pm(\cdot)$ is strongly analytic,  and $E(\cdot)$ analytic on  
$\dz\equiv \{(e, \es )\in\RR^2 | |e|<e_0, |\es|<\pr\}\subset\dgg$.
\el
\proof 
We write $K\kp$ as 
$K\kp=K_0+K_{{\rm I}}$, 
where 
\eqn 
&& K_0\equiv \half(\es n_z-\pf)^2+\hf,\\
&& K_{{\rm I}}\equiv -e(\es n_z-\pf) \cdot A +\frac{e^2}{2} \wick{AA}
 -\frac{e}{2} \s B.
\enn
For $(e, \es )$  with both of $|e|$ and $|\es|$   sufficiently small, 
we  see that 
\eq{anl2}
\|K_{{\rm I}}\Psi\|_\hk \leq a \|K_0 \Psi\|_\hk +b\|\Psi\|_\hk
\en 
for $\Psi\in D(K_0)$  with $a<1$ and $b\geq 0$.  
It can be seen by  
\kak{anl2},  \cite[p.16 Lemma]{rs4} and \cite[Theorem XII.9]{rs4} 
that $K_\pm\kp$ is an analytic family in the sense of Kato.  
Since $E\kp$ is an isolated non-degenerate eigenvalue of 
$K_\pm\kp$ by Lemmas \ref{tani} and \ref{hio},  
${\kr}_\pm\kp$ 
is strongly analytic in $e$,  and $E\kp$ analytic in $e$ by 
\cite[Theorem XII.8]{rs4}. Analyticity for $\es$ is also proven in 
the similar manner to $e$. Then the proof is complete. 
\qed
\noindent 
Under the identification 
$\hhh\cong \hk\otimes \fk$, 
we define 
\eq{er}
{\gr}_\pm\kp\equiv {\kr}_\pm\kp\otimes\Omega_\kappa,
\en 
where 
$\Omega_\kappa$ denotes the Fock vacuum in $\fk$. \\
{\it Proof of Lemma \ref{anal}}

Since $\he-E\kp=(K\kp-E\kp)\otimes 1_{\fk}+1_\hk\otimes 
(\hf\lceil_{\fk})$ and 
the dimension of ${\rm Ker} (K\kp-E\kp)$ is two and 
that of ${\rm Ker} (\hf\lceil_{\fff_0})$ is one, 
it follows that the dimension of ${\rm Ker} (\he-E\kp)$ is two. 
The analytic property of ${\gr}_\pm(\cdot)$ follows from Lemma \ref{ana} and \kak{er}. 
Thus the lemma is proven. 
\qed
Let $\es=0$ and set 
$$ 
\gr(e,0)=\sum_{n=0}^\infty \frac{e^n}{n!}\g n.$$
\bl{anal2}
There exists a strongly analytic ground state $\gr(e, 0)$ such that 
\begin{eqnarray} 
&& \label{koul} \g 0=\varphi_+,\\ 
&& \label{ana1} (\varphi_\pm, \g n)_\hhh=0,\ \ \ n=1,2,3.
\end{eqnarray} 
\el
\proof
Let $\gr(e,0)={\kr}_+(e,0)\otimes\Omega_\kappa$. 
Since $\varphi_-\in\hhh(-\han)$ and 
$\gr(e,0)\in\hhh(\han)$, \kak{koul} and 
\kak{ana1} for $\varphi_-$ follow. 
We shall prove \kak{ana1} for $\varphi_+$.  
Let $\d \fg(e)=\sum_{n=0}^\infty \frac{a_n}{n!}e^n$ be an analytic 
function and set 
$\d \fg(e)\gr(e,0)=\sum_{n=0}^\infty \frac{e^n}{n!}\tilde{\varphi}_{(n)}$. 
We can adjust  $a_n$ 
to satisfy  $(\varphi_\pm, \tilde{\varphi}_{(n)})_\hhh=0$ for $n=1,2,3$. 
Thus redefining $\gr(e,0)$ by $\fg(e)\gr(e,0)$, we have \kak{ana1}. 
\qed

%%%%%%%%%%%%%%%%%%%%%%%%%%%%%%%%%%%%%%%%%%%%%%%%%%%%%
\subsection{Main theorem}
%%%%%%%%%%%%%%%%%%%%%%%%%%%%%%%%%%%%%%%%%%%%%%%%%%%%%
By Lemma \ref{anal}, $m/\mass$ is analytic in $e$ near $e=0$. 
Let 
$${\mass}/{m} =\sum_{n=0}^\infty a_n(\La/m) e^{2n}.$$
\bt{main}
There exist constants $b_1>0,b_2>0, c_1>0$ and $c_2>0$ such that 
\eqnn 
&& 
\label{main1}
b_1\leq \liml \frac{a_1(\La/m)}{\log(\La/m)}\leq b_2, \\
&& 
\label{main2}
-c_1\leq \liml \frac{a_2(\La/m)}{(\La/m)^2}\leq -c_2.
\ennn
\et
\begin{remark}
This theorem implies that $a_{2}(\Lambda/m)<0$ for large $\Lambda$. 
Then the reader may think that this  contradicts  
to the fact that $\mass>m$ for $e^{2}>0$. However, 
the expansion is asymptotic and the leading term $a_{1}(\Lambda/m)$
is positive, then this negative sign is not a contradiction.
\end{remark}
To prove this theorem we derive exact forms of $a_2(\La/m)$ in the next section. 
%%%%%%%%%%%%%%%%%%%%%%%%%%%%%%%%%%%%%%%%%%%%%%%%%%%%%
\subsection{Expansions}
%%%%%%%%%%%%%%%%%%%%%%%%%%%%%%%%%%%%%%%%%%%%%%%%%%%%%
In what follows we assume that ground state 
$\d \gr=\sum_{n=0}^\infty\frac{e^n}{n!}\g n$  satisfies  
\kak{koul} and \kak{ana1}. Let us define $H$, $E$ and $\gr$ by 
\eqn 
&& H\equiv H(e,0)=
H_0+e\hi+\frac{e^2}{2}\hii, \\
&& E\equiv E(e,0)=\is(H),\\
&& \gr\equiv \gr(e, 0)=\sum_{n=0}^\infty\frac{e^n}{n!}\g n, 
\enn 
where 
\eqn 
&&H_0\equiv \hf+\half \pf^2,\\
&&\hi\equiv \hi^{(1)}+\hi^{(2)}, \ \ \ 
\hi^{(1)}=A \pf,\ \ \ \hi^{(2)}\equiv -\half \s B ,\\
&&\hii\equiv   \wick{A A}=A^+A^++2A^+A^-+A^-A^-. 
\enn
Here we put 
$$ 
A^+\equiv  
\frac{1}{\sqrt2}\jj \int \frac{\vpm(k)}{\sqrt{\omega(k)}}e(k,j) \add(k,j) dk, 
 \ \ 
A^-\equiv  
\frac{1}{\sqrt2}\jj \int \frac{\vpm(k)}{\sqrt{\omega(k)}}e(k,j) a(k,j) dk.
$$
\bl{exp}
It follows that 
$ \e 0=\e {2m+1}=0$, $m\geq 0$, 
and 
\eq{p2} 
\half \e 2=
(\g 0, \hi^{(2)}\g 1)_\hhh=-(\g 0, \bs\z\bs\g 0)_\hhh\not=0.
\en 
Moreover 
\eqnn
&&\g 0=\varphi_+,\\
&& \label{koma5} \g 1=-\z \hi^{(2)}\g 0, \\
&& \label{koma6} \g 2=\z(-\hii\g 0+2\hi\z\hi^{(2)}\g 0+\e 2 \g 0),\\
&& \label{koma7} \g 3=3\z(-\hii\g 1 -\hi\g 2+\e 2\g 1).
\ennn 
In particular 
$\g 1\in \cf 1$, $\g 2\in \cf 2$ and $\g 3\in \cf 1\oplus  \cf 3$. 
\el
\begin{remark}
Although  $\z$ is an unbounded operator, 
\kak{koma5}-\kak{koma7} are well defined by the fact 
that 
\eq{lieb}
D(\z)\supset 
\lkk
\{\left. \Psi^{(n)}\}_{n=0}^\infty \in\ffff \right |   \Psi^{(0)}=0,
{\rm supp}_{k \in \RR^{3n}}
 \Psi^{(n)}(k, j)\not\ni
\{0\},j\in\{1,2\}^n,n\geq 1\rkk. 
\en
\end{remark}
\proof 
It is obvious that
$\e 0=0$, 
and by the unitary equivalence between $H$ and $H$ with $e$ replaced by $-e$,  
$\e {2m+1}=0$ follows. 
Let $\gr'$ and $E'$ be $\st d\gr/de$ and $dE/de$, respectively, 
and  $\Phi\in D(H)$. Then 
\eq{both}
(H\Phi, \gr)_\hhh=E(\Phi, \gr)_\hhh.
\en 
Take derivatives in $e$ on the both sides of \kak{both}.
Then we have 
\eqnn
&&\label{a1}
(H'\Phi, \gr)_\hhh+(H\Phi, \gr')_\hhh=E'(\Phi, \gr')_\hhh+E(\Phi, \gr')_\hhh,\\
&&
(H''\Phi, \gr)_\hhh+2(H'\Phi, \gr')_\hhh+(H\Phi, \gr'')_\hhh\non\\
&&\label{a2}\hspace{2cm}
=E''(\Phi, \gr)_\hhh+2E'(\Phi, \gr')_\hhh+E(\Phi, \gr'')_\hhh,\\
&&
(H'''\Phi, \gr)_\hhh+3(H''\Phi, \gr')_\hhh+3(H'\Phi, \gr'')_\hhh+(H\Phi, \gr''')_\hhh\non\\
&&\label{a3}\hspace{2cm}
=E'''(\Phi, \gr)_\hhh+3E''(\Phi, \gr')_\hhh+3E'(\Phi, \gr'')_\hhh+E(\Phi, \gr''')_\hhh,
\ennn
where $H'=\hi+e\hii$ and $H''=\hii$. 
By \kak{a1} we see that $\gr'\in D(H)$ with 
\eq{11}
H'\gr+H\gr'=E'\gr+E\gr'.
\en
By \kak{a2} and \kak{a3}, we also see that 
$\gr''\in D(H)$ and $\gr'''\in D(H)$ with 
\eqnn 
&&\label {12}
H''\gr+2H'\gr'+H\gr''=E''\gr+2E'\gr'+E\gr'',\\
&&\label {13}
H^{'''}\gr+3H''\gr'+3H'\gr''+H\gr^{'''}=E^{'''}\gr+3E''\gr'+3E'\gr''+E\gr^{'''}
\ennn
From \kak{12} it follows that
\eq{m6}
(\gr, H''\gr)_\hhh+(\gr,2H'\gr')_\hhh+(\gr,H\gr'')_\hhh=E''(\gr,\gr)_\hhh+
(\gr,2E'\gr')_\hhh+(\gr,E\gr'')_\hhh.
\en
Put $e=0$ in \kak{m6}.
Then
\eq{39}
\hspace{-1cm}
(\g 0 , \hii\g 0 )_\hhh+(\g 0 , 2\hi\g 1 )_\hhh+
(\g 0 , H_0\g 2)_\hhh=\e 2(\g 0 ,\g 0 )_\hhh.
\en
Since the left-hand side of \kak{39} equals to 
$2(\g 0, \hi^{(2)}\g 1)_\hhh$, 
it follows that 
$\e 2=2(\g 0, \hi^{(2)}\g 1)_\hhh$. 
By \kak{11}, \kak{12} and \kak{13} with $e=0$,
we have
\eqn 
&& \hi^{(2)}\g 0 +H_0\g 1=0,\\
&&\hii\g 0 +2\hi\g 1 +H_0\g 2 = \e 2 \g 0 ,\\
&& 3\hi\g 2 +3 \hii \g 1 +H_0\g 3 =3 \e 2 \g 0.
\enn
Then 
\eqn  
&&\g 1= -\z \hi^{(2)}\g 0 + a\varphi_++b\varphi_-,\\
&& \g 2= -\z (\hii\g 0 +2\hi\g 1- \e 2 \g 0) 
 + a'\varphi_++b'\varphi_-,\\
&& 
\g 3= 
\z (-
3\hi\g 2 -3 \hii \g 1 +3 \e 2 \g 0) 
+ a''\varphi_++b''\varphi_-
\enn 
with some constants $a$, $b$, $a'$,  $b'$, $a''$ and $b''$. 
Since we see that 
$-\z \hi^{(2)}\g 0\perp \fff^{(0)}$, 
$-\z (\hii\g 0 +2\hi\g 1-\e 2 \g 0) \perp \fff^{(0)}$ and 
$\z (-3\hi\g 2 -3 \hii \g 1 +3 \e 2 \g 0) \perp \fff^{(0)}$, 
it follows that
$a=b=a'=b'=a''=b''=0$ by  \kak{ana1}. 
Then 
\kak{koma5}, \kak{koma6} and 
\kak{koma7} follow. 
Then the lemma is proven.
\qed
\bl{hi}
Let  $\grr'=\st \partial \gr(e, \es)/\partial \es\lceil_{\es=0}$. 
Then $(\pf+eA)_{\mu=3}\gr\in D((H-E)\f)$ with 
\eq{226}
\gr'=(H-E)\f(\pf+eA)_{\mu=3}\gr,
\en 
and 
\eq{24}
\frac{m}{\mass}=
1-2  \frac{(\gr, (\pf+eA)_3{\gr}')_\hhh}{(\gr,\gr)_\hhh}.
\en 
\el
\proof 
Since 
$\he\gr(e, \es)=\Ee\gr(e, \es)$, 
in the similar manner as \kak{11}-\kak{13}, we have 
\eqnn
&&\label{l1}
H'(e, \es)\gr(e, \es)+\he\grr'(e, \es)=E'(e, \es)\gr(e, \es)+\Ee\grr'(e, \es),\\
&&
H''(e, \es)\gr(e, \es)+2H'(e, \es)\grr'(e, \es)+\he\grr''(e, \es)\non\\
&&\hspace{3cm} \label{l2} 
=E''(e, \es)\gr(e, \es)+2E'(e, \es)\grr'(e, \es)+E(e, \es)\grr''(e, \es)
\ennn
where $\grr'\kp=\st\partial\gr\kp/\partial \es$, 
$H'(e, \es)=-(\pf+eA)_3$ and $H''(e, \es)=0$. 
Putting $\es=0$ in \kak{l1} and \kak{l2}, we see that 
\eqnn
&&\label{l3}(H-E)\grr'=(\pf+eA)_{\mu=3}\gr,\\
&&\label{l4} E''\gr=(H-E)\grr''+\gr-2(\pf+eA)_3\grr'. 
\ennn
Since ${m}/{\mass}= E''$ by \kak{01}, 
we have 
$$\frac{m}{\mass}= \frac{(\gr, E''\gr)_\hhh}{(\gr,\gr)_\hhh}.$$ 
Thus the lemma follows from \kak{l3} and \kak{l4}. 
\qed
\begin{remark}
\label{sta}
Combining \kak{24} with \kak{226} we have 
\eq{255}
\frac{m}{\mass}=
1-2  \frac{((\pf+eA)_3 \gr, (H-E)\f (\pf+eA)_3\gr)_\hhh}{(\gr,\gr)_\hhh}.
\en 
Or, by a symmetry we can also express $m/\mass$ 
such as 
\eq{25}
\frac{m}{\mass}=
1-\frac{2}{3}\mmm \frac{((\pf+eA)_\mu \gr, 
           (H-E)\f (\pf+eA)_\mu\gr)_\hhh}{(\gr,\gr)_\hhh}.
\en 
\end{remark}
\bc{po}
The effective mass satisfies that  $\mass>  m$ for $e\not=0$ but 
$|e|$ is sufficiently small. 
\ec
\proof 
Since 
$((\pf+eA)_\mu \gr, (H-E)\f (\pf+eA)_\mu\gr)_\hhh\geq0$, 
it follows that $\mass\geq m$  from \kak{25}. For $e=0$, $\mass=m$. 
Since $\mass$ is analytic in $e$, the corollary follows. 
\qed
\bt{3}
The effective mass  is expanded as 
\eq{haha2}
\frac{m}{\mass}=1-\frac{2}{3}c_1(\La/m)e^2-\frac{2}{3}c_2(\La/m) e^4+{\mathcal O}(e^6),
\en 
or 
\eq{haha3}
\frac{\mass}{m}=1+\frac{2}{3}c_1(\La/m)e^2+\lk 
\frac{2}{3}c_2(\La/m) +\lk\frac{2}{3}\rk^2 c_1(\La/m)^2\rk e^4+{\mathcal O}(e^6), 
\en 
where 
\eqnn
&&\hspace{-0.5cm}
c_1(\Lambda/m)\equiv \mmm (\G 1, \wh 0 \G1)_\hhh ,\nonumber\\
&&\hspace{-0.5cm}
 c_2(\Lambda/m) \equiv  
\mmm \lkk (\G 1, \wh 2 \G1)_\hhh-(\G1, \wh0\G1)_\hhh (\g 1,\g 1)_\hhh+
2\Re (\G2,\wh1\G1)_\hhh\right.\nonumber \\ 
&& \hspace{6cm} \label{koma1}
\left.+(\G2,\wh0\G2)_\hhh+2\Re (\G3,\wh0\G1)_\hhh\rkk. 
\ennn
Here 
$$\G n\equiv \frac{1}{(n-1)!} A_\mu\g {n-1}+\frac{1}{n!} \pf_\mu\g n,\ \ \ n=1,2,3,\ \ \ 
\mu=1,2,3,$$
$$ 
\wh 0=\z, \ 
\wh 1 =-\z \hi\z,\ 
\wh 2=-\half \z \hii \z+\z(\hi\z\hi-\EE)\z. 
$$
\et 
\proof 
We have 
\eq{haha1}
\frac{1}{(\gr,\gr)_\hhh}= 1-e^2 (\g 1, \g 1)_\hhh -e^4 (
  \frac{1}{4} (\g 2, \g 2)_\hhh
+ \frac{1}{6}
(\g 1, \g 3)_\hhh) +{\mathcal O}(e^6)
\en
and 
\eq{koma8}
(\pf+e A)_\mu\gr=  e\G 1+e^2\G 2+e^3 \G 3+{\mathcal O}(e^4).
\en 
Note also that 
$ \G 1\in \cf 1$, $\G 2\in \cf 2$ and $\G 3\in \cf 1\oplus \cf 3$. 
Let us set 
\eq{ha2}
\grr'\equiv \sum_{n=0}^\infty 
   \frac{e^n}{n!}\ppp n,\ \ \ \mu=1,2,3.
\en 
Substituting \kak{haha1}, \kak{koma8} and \kak{ha2}  into \kak{24}, we have 
\eqnn
&& 
\hspace{-1cm}\frac{m}{\mass}=
1-2 
(\Gt 1, \ppp 1)_\hhh e^2\non \\
&& \label{coin}
%\hspace{1cm}
-2 
\lkk 
\frac{1}{6} (\Gt1, \ppp 3)_\hhh+\half (\Gt 2, \ppp 2)_\hhh+
\frac{1}{6} (\Gt 3, \ppp 1)_\hhh\rkk
e^4+{\mathcal O}(e^6).
\ennn 
Then we shall see the explicit form of $\ppp n$ below.
The identity 
\eq{sou}
((H-E)\Phi, \gr')_\hhh=((\pf+eA)_3\Phi, \gr)_\hhh,\ \ \ 
\Phi\in D(H),
\en is derived from \kak{l3}.  
Putting  $e=0$ in \kak{sou}, we have 
$ H_0\ppp 0=0$, 
which implies that 
$ 
\ppp 0=\ab 0
$ 
with some constants $a_0$ and $b_0$. 
We differentiate both sides of \kak{sou} 
at $e=0$. 
Then we have 
\eqn
&&\label{o1} \hi\ppp 0+H_0\ppp 1=A_3\g 0+\pf_3\g 1,\\
&&\label{o22} (\hii-\e 2)\ppp 0 +2\hi\ppp 1+H_0\ppp 2=2A_3\g 1+\pf_3\g 2,\\
&&\label{o11} 3(\hii-\e 2)\ppp 1 +3\hi\ppp 2+H_0\ppp 3=3A_3\g 2+\pf_3\g 3, 
\enn  
from which it follows that 
\eqn
&&\label{e1}
\ppp 1 =\z (A_3\g 0+\pf_3\g 1-\hi\ppp 0)+\ab 1,\\
&&\label{e2}
\ppp 2=
\z(2A_3\g 1+\pf_3 \g 2-2\hi\ppp 1)+\ab 2,\\
&&\label{e3}
\ppp 3=
\z(3A_3\g 2+\pf_3 \g 3-3\hi\ppp 2-3(\hii-\e 2)\ppp 1)+\ab 3, 
\enn
where $a_j,b_j$, $j=1,2,3$,  are some constants. 
Then 
it is obtained that 
\eqn 
&&\ppp 1=\z\Gt 1+R_1,\\
&&\ppp 2=\z(2\Gt 2-2\hi \z\Gt 1)+R_2,\\
&&\ppp 3=\z(6\Gt 3-6\hi\z\Gt 2+6\hi\z\hi\z\Gt 1-3(\hii-\e 2)\Gt 1)+R_3.
\enn
Here 
\eqn
&& R_1\equiv -\z\hi(\ab 0)+\ab 1,\\
&& R_2\equiv 2\z\hi\z\hi(\ab 0)-2\z\hi (\ab 1)+\ab 2,\\
&& R_3\equiv 
-6\z\hi\z\hi\z\hi(\ab 0)+3(\hii-\e 2)\z\hi(\ab 0)\\
&&\hspace{2cm} 
 +6\z\hi\z\hi (\ab 1)-3\z(\hii-\e 2)(\ab 1)\\
&&\hspace{2cm} 
-3\z\hi (\ab 2)+\ab 3.  
\enn
It is directly seen that the contributions 
of constants $a_0,b_0,...,a_3,b_3$ to \kak{coin} are 
$$ 
(\Gt 1, R_1)_\hhh =
(\Gt 1, R_3)_\hhh =(\Gt 2, R_2)_\hhh =(\Gt 3, R_1)_\hhh=0.
$$
Then \kak{coin} coincides with \kak{haha2} by a symmetry. 
Hence the lemma follows. 
\qed
We shall compute $c_2(\La/m)$. 
We set 
$$ 
\wh 0 \equiv  \q 1, \ \
 \wh 1 \equiv \q 2+\q 3,\ \
\wh 2 \equiv  
\q 4+\q 5+\q 6+\q 7+\q 8,
$$ 
where we put
\begin{eqnarray*}
&& 
\q 1 = \z,\ \ \ 
\q 2 =-\z\pfa\z,\ \ \ 
\q 3 = -\z\bs\z, \\ 
&& 
\q 4 = \half\z(-\hii)\z,\ \ \ 
\q 5 = \z\pfa\z\pfa\z, \\ 
&&
\q 6 =  \z\bs\z\pfa\z,\ \ \ 
\q 7 =  {\q 6}^{*} = \z\pfa\z\bs\z,\\
&& 
\q 8 = \z \left\{ \bs\z\bs -\EE \right \}\z.
\end{eqnarray*}
We list groups of vectors by  which $\gr$ is expanded:
\begin{eqnarray*}
\g 1 &=&-\z \bs \g 0, \\
\g 2 &=&\z(-\hii)\g 0 +2\z \pfa \z \bs \g 0 \\
     & &+2 \z\left\{\bs \z \bs -\EE  \right \}  \g 0,\\
\g 3 &=&-3\z(-\hii)\z\bs\g 0+3\z \pfa \z (-\hii)\g 0\\
     & &+6\z \pfa \z \pfa \z \bs \g 0\\
     & &+6\z \pfa \z\left\{\bs \z \bs-\EE  \right\} \g 0\\
     & &+3\z \bs\z (-\hii) \g 0+6\z\bs\z \pfa \z \bs\g 0\\
     & &+6\z\bs\z\left\{\bs\z\bs-\EE \right\}\g 0.\\
\end{eqnarray*}
From the above expressions of $\g 1, \g 2, \g 3$,  it follows 
that for $\mu=1,2,3$, 
$$ 
\G 1 
\equiv \pp 1+\pp 2, \ \
\G 2 
\equiv \sum_{i=2}^6 \pp i, \ \
\G 3 
\equiv \sum_{i=7}^{16} \pp i,
$$ 
where 
\eqn 
&&\pp 1=A_\mu^+\g 0\\
&&\pp 2=\half\pf_\mu \z \sbb \g 0\\
&&\pp 3=\half A_\mu\z\sbb \g 0\\
&&\pp 4=-\half \pf_\mu \z {A^+\! A^+} \g 0\\
&&\pp 5=-\half \pf_\mu \z \pfa\z \sbb\g 0\\
&&\pp 6=\pf_\mu\z\lkk \bs \z\bs-\EE\rkk \g 0\\
&& \ \ \ \ \ \ 
=\frac{1}{4}\pf_\mu \z \sbb\z \sbb\g 0\\
&&\pp 7=-\half A_\mu\z{A^+\! A^+}\g 0\\
&&\pp 8=-\half A_\mu \z \pfa \z \sbb\g 0\\
&&\pp 9= \half A_\mu 2\z \lkk \bs \z\bs-\EE\rkk  \g 0\\
&& \ \ \ \ \ \  =\frac{1}{4}A_\mu\z\sbb\z\sbb\g 0\\
&&\pp {10}=-\frac{1}{4}\pf_\mu\z\wick{AA}\z\sbb\g 0\\
&&\pp {11}= -\half \pf_\mu \z \pfa \z {A^+\! A^+}\g 0\\
&&\pp {12}=-\half \pf_\mu\z\pfa\z\pfa\z\sbb\g 0\\
&&\pp {13}= 
\frac{1}{6}\pf_\mu 3 \z \pfa 2 \z \lkk \bs \z \bs-\EE\rkk  \g 0\\
&& \ \ \ \ \ \  \ =\frac{1}{4}\pf_\mu \z\pfa\z\sbb\z\sbb\g 0\\
&&\pp {14}= \frac{1}{4}\pf_\mu\z\s  B\z{A^+\! A^+}\g 0\\
&&\pp {15}= \frac{1}{4}\pf_\mu\z\s  B\z\pfa\z\sbb\g 0\\
&&\pp {16}=
\frac{1}{6}\pf_\mu 3 \z \bs 2\z \lkk \bs \z \bs-\EE\rkk  \g 0\\
&& \ \ \ \ \ \ \  =-\frac{1}{8}\pf_\mu\z\s B \z\sbb\z\sbb\g 0
\enn 
\begin{remark}
\label{hos}
We give a remark on terms $\pp 6, \pp 9, \pp{13}, \pp{16}$. 
Since by \kak{p2}
$$(\g 0, \lkk \bs \z \bs -\EE\rkk\g 0)_\hhh=0,$$
we see that 
$\d \lkk \bs \z \bs -\EE\rkk\g 0$ is perpendicular to $\fff^{(0)}$.
Thus we have 
$$\lkk \bs \z \bs -\EE\rkk\g 0= 
\frac{1}{4}\s  B^+\z \s  B^+\g0\in \fff^{(2)}.$$
\end{remark}
Substituting  $H_1,...,H_8$ and $\pp 1, ..., \pp {16}$ 
  into \kak{koma1}, we have the  lemma below: 
\bl{cm}
The coefficient $c_2(\La/m)$ is given by 
\eqnn 
&&\hspace{-1cm}
c_2(\La/m)
=
\mmm \lkk 
(\sum_{i=1}^2\pp i, \sum_{\ell=4}^8 \q \ell \sum_{i=1}^2\pp i)_\hhh\right.
+
(\sum_{i=1}^2\pp i, \q 1 \sum_{i=1}^2\pp i)_\hhh\non\\
&&\hspace{-1cm}\label{76}
(\sum_{i=3}^6 \pp i, (\q 2+\q 3) \sum_{i=1}^2\pp i)_\hhh
+
(\sum_{i=3}^6 \pp i, \q 1 \sum_{i=3}^6 \pp i)_\hhh
\left.+
(\sum_{i=7}^{16} \pp i, \q 1 \sum_{i=1}^2\pp i)_\hhh\rkk.
\ennn 
\el
From \kak{76} we see that $c_2(\La/m)$ is decomposed into 76 terms. 

%%%%%%%%%%%%%%%%%%%%%%%%%%%%%%%%%%%%%%%%%%%%%%%%%%%%
\subsection{Classification of the Divergent Terms}
%%%%%%%%%%%%%%%%%%%%%%%%%%%%%%%%%%%%%%%%%%%%%%%%%%%%
Though we have 76 terms of the form 
$(\pp \ell, H_{m}\pp n)_\hhh$ in $c_{2}(\Lambda/m)$, 
not all of them are important. Namely $(\pp \ell, H_{m} \pp n)_\hhh$
and its complex conjugate $(\pp n, H_{m}^{*} \pp \ell)_\hhh$ appear
and  we take the real part of them.  
\bl{ito}
It follows that  $\d \sum_{{\rm \stackrel{(\pp \ell, H_m\pp n)_\hhh
\mbox{ contains}}{\mbox{ odd number of } \s  B}}}
(\pp \ell, H_m\pp n)_\hhh=0$. 
\el
\proof
Let  $\gamma\equiv  C \sigma_{3}$,
where $C$ stands for a complex-conjugacy operator on $\fff$:
$C^{2}=1$, $C J=-J C$, $J=\sqrt{-1}$. We choose $\varphi_{(0)}$ so that 
$\gamma \varphi_{(0)}=\varphi_{(0)}$. Then 
$\gamma$ is an anti-unitary operator such that 
$\gamma \sigma_\mu \gamma^\ast=\sigma_\mu$, 
$\gamma A_{\mu} \gamma^\ast=A_{\mu}$,
$\gamma B_{\mu} \gamma^\ast=-B_{\mu}$
and 
$(\psi,\phi)_{\fff}=(\gamma \phi,\gamma \psi)_{\fff}$. Thus 
terms containing odd number of $\sigma B$ are purely  imaginary.
Then the sum of contributions of terms containing 
odd number of $\sigma B$ is zero. The proof is complete. 
\qed
By this lemma 
it is enough to consider terms containing even number of $\sigma B$'s. 
Our results are summarized on the lists for $c_{2}(\Lambda/m)$ 
which consist of five groups. 
The columns of $\sigma B$ (resp. $\pf$, $\z$) 
mean the number of $\sigma B$ (resp. $\pf$, $\z$) in term 
$(\pp \ell,  H_m\pp n)$. In Tables \ref{op1}-\ref{op5}, we exhibit
only the terms containing even  number of $\s  B$.  On these lists,
order means an upper or a lower bound for the leading divergent 
term of the expectation values $(\Phi_{(\ell)},H_{(m)}\Phi_{(n)})$ 
listed on the row.  Note that they are one of 
$[\log(\La/m)]^{2}$, $\sqrt{\Lambda/m}$ and $(\Lambda/m)^{2}$. 

\begin{table}[p]
\begin{center}
\arrayrulewidth=1pt
\def\arraystretch{1.2}
\begin{tabular}{|c|c|c|c|c|c|}
\hline
No.   &  Term  &  $\sigma B$ &   $\pf$ &   $\z$ &  Order \\
\hline 
(1)   &  $-\RRR 1 1 1 (\g 1, \g 1)_\hhh$ &2&0& 3&  $\l 2$ \\
\hline
(2)   &  $-\RRR 2 1 2 (\g 1, \g 1)_\hhh$ &4&2& 5&  $\l 2$ \\
\hline
\end{tabular}
\end{center}
\caption{
    $(\sum_{i=1}^2\pp i,  H_1 \sum_{i=1}^2\pp i)_\hhh$}
\label{op1}
\end{table}
\begin{table}[p]
\begin{center}
\arrayrulewidth=1pt
\def\arraystretch{1.2}
\begin{tabular}{|c|c|c|c|c|c|}
\hline
No.   &  Term  &  $\sigma B$ &   $\pf$ &   $\z$ &  Order \\
\hline 
(1) & $\RRR 3 1 3 $ &2&0& 3&  $[\log(\La/m)]^2$  \\
\hline 
(2) & $\RRR 5 1 3 $ &2&2& 4&   $\l 2$ \\
\hline 
(3) & $\RRR 3 1 5$ &2&2 &4 &  $\l 2$ \\
\hline
(4) & $\RRR 4 1 4 $ &0&2& 3&  $+\sqrt{\La/m}$\\
\hline  
(5) & $\RRR 6 1 4 $ &2&2& 4&  $-\sqrt{\La/m}$ \\
\hline
(6) & $\RRR 4 1 6$ &2&2& 4&  $-\sqrt{\La/m}$\\
\hline
(7) & $\RRR 5 1 5$ &2&4& 5 &  $+\sqrt{\La/m}$ \\
\hline 
(8) & $\RRR 6 1 6$ &4&2& 5 &  $\l 2$ \\
\hline
\end{tabular}
\end{center}
\caption{
     $
          (\sum_{i=3}^6\pp i, 
          \q 1 \sum_{i=3}^6\pp i )_\hhh$ }
\label{op2}
\end{table}

\begin{table}[p]
\begin{center}
\arrayrulewidth=1pt
\def\arraystretch{1.2}
\begin{tabular}{|c|c|c|c|c|c|}
\hline
No.   &  Term  &  $\sigma B$ &   $\pf$ &   $\z$ &  Order \\
\hline 
(1) & $\RRR 1  4 1 $ &0&0& 2&  $[\log(\La/m)]^2$\\
\hline
(2) & $\RRR 2 4 2 $ &2&2& 4& $\l 2$ \\
\hline 
(3) & $\RRR 1 5 1 $ &0&2& 3& $\log(\La/m)$\\
\hline 
(4) & $\RRR 2 5 2 $ &2&4& 5& $\l 2$ \\
\hline
(5) & $\RRR 2 6 1 $ &2&2& 4& $\l 2$ \\
\hline
(6) & $\RRR 1 6 2 $ &2&2& 4&  $\l 2$\\
\hline
(7) &$\RRR 2 7 1  $ &2&2& 4&  $\l 2$ \\
\hline
(8) & $\RRR 1 7 2 $ &2&2& 4&  $\l 2$ \\
\hline
(9) &$\RRR 1 8 1  $ &2&0& 3&  $\l 2$ \\
\hline
(10) & $\RRR 2 8 2$ &4&2& 5&  $\l 2$ \\
\hline 
\end{tabular}
\end{center}
\caption{$(\sum_{i=1}^2\pp i,
      \sum_{\ell=4}^{8} \q \ell \sum_{i=1}^2\pp i)_\hhh$ }
\label{op3}
\end{table}

\begin{table}[p]
\begin{center}
\arrayrulewidth=1pt
\def\arraystretch{1.2}
\begin{tabular}{|c|c|c|c|c|c|}
\hline
No. & Term  &  $\sigma B$ & $\pf$ & $\z$  &    Order \\
\hline
(1) & $\RRR 4 2 1 $ &0&2& 3& $\log(\La/m)$ \\
\hline
(2) & $\RRR 6 2 1 $ &2&2& 4&   $\l 2$ \\
\hline 
(3) & $\RRR 3 2 2 $ &2&2& 4&  $\l 2$ \\
\hline
(4) & $\RRR 5 2 2 $ &2&4& 5&  $\sqrt{\La/m}$\\
\hline
(5) & $\RRR 3 3 1 $ &2&0& 3&  $\l 2$\\
\hline
(6) & $\RRR 5 3 1 $ &2&2& 4& $\l 2$ \\
\hline
(7) & $\RRR 4 3 2 $ &2&2& 4& $\sqrt{\La/m}$ \\
\hline
(8) & $\RRR 6 3 2 $ &4&2&  5&  $-\sqrt{\La/m}$ \\ 
\hline
\end{tabular}
\end{center}
\caption{ $ (\sum_{i=3}^{6}\pp i, (H_{2}+H_{3})(\sum_{i=1}^2\pp i))_\hhh$
}
\label{op4}
\end{table}

\begin{table}[p]
\begin{center}
\arrayrulewidth=1pt
\def\arraystretch{1.2}
\begin{tabular}{|c|c|c|c|c|c|}
\hline
No. & Term  &  $\sigma B$ & $\pf$ & $\z$  &    Order \\
\hline
(1) & $\RRR 7 1 1$&0&0&  2&  $[\log(\La/m)]^2$ \\
\hline
(2) & $\RRR 9 1 1$&2&0& 3&$\l 2$ \\
\hline
(3) & $\RRR {11}1 1$ &0&2& 3& $=0$ \\
\hline
(4) & $\RRR {13} 1 1$ &2&2& 4& $=0$ \\
\hline
(5) & $\RRR {15} 1 1 $ &2&2& 4& $=0$ \\
\hline
(6) & $\RRR 8 1 2    $ &2&2& 4& $\l 2$ \\
\hline
(7) & $\RRR {10} 1 2 $ &2&2& 4&   $=0 $\\
\hline
(8) & $\RRR {12} 1 2 $ &2&4& 5& $\l 2$ \\
\hline
(9) & $\RRR {14} 1 2 $ &2&2& 4& $\l 2$ \\
\hline
(10) & $\RRR {16} 1 2 $& 4&2& 5& $-(\Lambda/m)^{2}$ \\
\hline
\end{tabular}
\end{center}
\caption{      $  (\sum_{i=7}^{16} \pp i,    H_1\sum_{i=1}^2\pp i)_\hhh$
}
\label{op5}
\end{table}

%%%%%%%%%%%%%%%%%%%%%%%%%%%%%%%%%%%%%%%%%%%%%%%%%%%%%%%%%%%%%%%%%%%%%
\section{Estimates of 38 terms}
%%%%%%%%%%%%%%%%%%%%%%%%%%%%%%%%%%%%%%%%%%%%%%%%%%%%%%%%%%%%%%%%%%%%%
We classify these 38 terms into 3 types by the numbers of $\sigma B$'s.
Type I consists of terms which contain no $\sigma B$
and thus has been already calculated in \cite{hisp2}.
Type II consists of terms which contain two $\sigma B$'s  and 
two  $A_{\mu}$'s.  
Finally Type III consists of terms which contain four $\sigma B$'s only, 
and diverges most strongly.
%%%%%%%%%%%%%%%%%%%%%%%%%%%%%%%%%%%%%%%%%%%%%%%%%%%%%%%%%%%%%%%%%%%%%
\subsection{Terms with four $\sigma B$'s  or  ${1\over2}E_{(2)}$}
%%%%%%%%%%%%%%%%%%%%%%%%%%%%%%%%%%%%%%%%%%%%%%%%%%%%%%%%%%%%%%%%%%%%%
Though the conventional power-counting theorem says that these 38 
quantities diverge at most like $\l 2$ as $\La\rightarrow \infty$, 
this does not work in some cases. 
The first violation takes place in the region where two momenta
$k_{1}\in \BR$  and $k_{2}\in \BR$ of two photons have 
opposite directions and $|k_{1}+k_{2}|$ becomes small. Such a violation yields 
$\sqrt{\La/m}$ divergence, which is investigated in \cite{hisp3}.  
The second violation takes place when the integrand is far 
from symmetric; each term $\RRR i j k$,  roughly speaking, has 
the form 
$${\cal E}=\int_{\kappa/m\leq |k_1|,|k_2|\leq \La/m} 
        dk_1 dk_2 |k_{1}|^{-a}|k_{2}|^{-b},$$
where $1\leq a$, $1\leq b$,  $a+b=6$ 
(see Appendix). Then 
$$c_{1}\l 2 \leq {\cal E}  \leq c_{2}(\Lambda/m)^{2}$$
depending on $a$ and $b$, with some constants $c_1$ and  $c_2$. 
Thus we restrict ourselves to the most singular and complicated 
contributions  of order $e^{4}$,  which come from the terms of 
type III and $\EEE$.  See Table ~\ref{iii}. 
Note that $\EEE\approx - (\La/m)^2$ as $\La\rightarrow\infty$ 
which may be, however, regarded as a subtraction needed to {\it Wick order }
$\bs  \z  \bs$. 
Actually all the term $\pp i$ has no $\EEE$, 
but only $H_{8}$ contains it. 
\begin{table}[ht]
\label{tt}
\begin{center}
\arrayrulewidth=1pt
\def\arraystretch{1.2}
\begin{tabular}{|c|c|c|c|}
\hline 
 Term &   $\sigma B$&  $\pf$ &  $\z$ \\
\hline 
 $\ce_0=\RRR 1 8 1 $ & 2 & 0& 3 \\
\hline
 $\ce_{1}=\RRR 6 3 2 $ & $4$ & $2$ & $5$ \\
\hline
 $\ce_{2}=\RRR 6 1  6 $ & $4$ & $2$ & $5$ \\
\hline
 $\ce_{3}=\RRR 2  8 2 $ & $4$ & $2$ & $5$  \\
\hline 
 $\ce_{4}=\RRR {16} 1 2 $ & $4$ & $2$ &$5$ \\
\hline
\end{tabular}
\caption{Terms with four $\s B$ and $\EEE$}
\label{iii}
\end{center}
\end{table}
%%%%%%%%%%%%%%%%%%%%%%%%%%%%%%%%%%%%%%%%%%%
\subsection{Lower and upper bounds}
%%%%%%%%%%%%%%%%%%%%%%%%%%%%%%%%%%%%%%%%%

\bl{divless}
The 35 terms except for $\ce_0$, $\ce_3$ and $\ce_4$ 
satisfy that 
$$\lim_{\La\rightarrow\infty} \frac{|\RRR i j k|}{ \sqrt{\La/m}}<\infty,
\ \ \ \mu=1,2,3,$$
where $(i,j,k)\not=(1, 8, 1), (2, 8, 2), (16, 1, 2)$.  
\el
Though this lemma can be proven in principle in the similar 
manner to \cite{hisp3}, we sketch the outline of our calculations 
in the Appendix for the sake of the reader who may not be familiar
with the perturbative calculations of the field theory.  We recommend 
the reader to read the Appendix in advance.
\begin{remark}
Each upper bound of $\RRR i j k$ is listed in Tables \ref{op1}-\ref{op5}. 
We note that $(\pp i,  H_{1} \pp 1)$, $i=11,13,15$  in Table 5 
are automatically zero since $\pf A \varphi_{(0)}=A \pf \varphi_{(0)}=0$.
Moreover $\RRR {10} 1 2 =0$ since the integrand is odd in momentum $k_{i}$.
\end{remark}

We write down  explicit forms of  $\ce_0,\ce_3$ and $\ce_4$ which
contain most singular divergence: 
\begin{eqnarray*}
\ce_{0}&=&\frac{1}{4}
         \mmm (\g 0 , A_{\mu}^{-}
               \z 
                \left\{ \sigma B \z 
                \sigma B -4 \EE \right\}  
\z  A_{\mu}^{+}\g 0 )_\hhh\\
&=&\frac{1}{4}
         \mmm (\g 0 , A_{\mu}^{-}
               \z 
              \sigma B^{-} \z \sigma B^{+}
                  \z  A_{\mu}^{+}\g 0 )_\hhh
         -\EE \AA,\\ 
\ce_{3}&=& \frac{1}{16}
         \mmm  (\g 0 ,
                \sigma B^{-} \z 
                \pf_\mu\z 
                \left\{ \sigma B \z 
                \sigma B -4 \EE \right\}
                 \z \pf_\mu
                \z  \sigma B^{+}\g 0 )_\hhh \\
 & = & \frac{1}{16}
         \mmm  (\g 0 ,
                \sigma B^{-}
                \pf_\mu (\z)^2 
               \sigma B^{-} \z \sigma B^{+} 
               (\z)^2 \pf_\mu
               \sigma B^{+}\g 0 )_\hhh 
             -\frac{1}{4}\EE \BB, \\
 \ce_{4}&=&  -\frac{1}{16}
                (\g 0 , \sigma B^{-} 
                (\z)^3 
                \pf^{2}  \sigma B^{-} 
                \z  
               \sigma B^{+} \z  \sigma B^{+}
                \g 0 )_\hhh.  
\end{eqnarray*}
Here we have used that the contribution from 
$\sigma B^{+}\z \sigma B^{-}$ in $\{\cdots\}$ in $\ce_1$ and $\ce_{3}$
vanishes, since 
$\s B^- \z A_\mu^+\g 0=0$ and $\s B^- (\z)^2\pf_\mu \s B^+\g 0=0$, 
and we put
\begin{eqnarray*}
-\frac{\e 2}{2}  &=& \frac{1}{4} \mmm  (\g 0 , \sigma B_{\mu}^{-}
           \z  \sigma B_{\mu}^{+}\g 0 )_\hhh,\\
\AA  &\equiv&   \mmm (\g 0 , A_{\mu}^{-}
               (\z)^2 A_{\mu}^{+}\g 0 )_\hhh, \\
\BB  &\equiv&   \mmm (\g 0 , \sigma B_{\mu}^{-}
               (\z)^4\pf^{2} \sigma B_{\mu}^{+}\g 0 )_\hhh.
\end{eqnarray*}
\bl{div}
Asymptotic behaviors of $\e 2$, $\AA$ and $\BB$ are 
(1) $\d \liml \left|\frac{\e 2}{(\La/m)^{2}} \right|<\infty$, 
(2) $\liml \AA<\infty$, 
(3) $\liml \BB<\infty$. 
\el
\proof 
Let us express the expectation value of $2\times 2$ matrix $O$ 
with respect to $\C 1 0$ by 
$$\AB O\equiv (\C 1 0, O\C 1 0).$$
Using the pull-through formula etc. (see Appendix), we see that 
\begin{eqnarray*}
\e 2 &=&  -{1\over2}\sum_{\ell=1,2}
          \int_{\kappa/m\leq |k|\leq\La/m} 
\frac{1}{(2\pi)^{3}2 \omega(k)}
          \frac{ \AB{(\sigma\cdot[k\times e(k,\ell)])^{2}}}
               {|k|^2/2 +|k|}dk  \nonumber \\
      &=&
-\frac{1}{8\pi^2}\int_{\kappa/m}^{\Lambda/m}\frac{r^2}{r+2} dr.
\end{eqnarray*}
Then (1) follows.  (2) and (3) follow from 
$$\liml |\AA|\leq {\rm const.} 
     \times \int_{\kappa/m}^\infty\frac{1}{r(r+2)^2} dr<\infty$$
and 
$$\liml |\BB|\leq  {\rm const.} \times  
     \int_{\kappa/m}^\infty\frac{1}{r(r+2)^2} dr<\infty.$$
Thus the lemma is proven. \qed
Using formulae
\begin{eqnarray}
&& \hspace{-0.5cm}
(\sigma\cdot p)(\sigma\cdot q)
   =   -(\sigma\cdot q)(\sigma\cdot p) + 2(p,q)  
   = (p,q)+i\sigma\cdot(p\times q), \quad
      (\sigma\cdot p)^{2}=|p|^{2},              \label{spin1}\\
&& \hspace{-0.5cm} 
(k_{1}\times e(k_1,\ell_1), k_{2}\times e(k_2,\ell_2))
   =(k_{1},k_{2})(e(k_1,\ell_1), e(k_2,\ell_2))-
      (k_{1},e(k_2,\ell_2))(k_{2},e(k_1,\ell_1)),\non \\               
&& \ \label{pol1}\\ 
&& \hspace{-0.5cm} 
 \sum_{\ell_2=1,2}(k_1,e(k_2,\ell_2))^{2}
  = |k_1|^{2}-(k_1,\hat{k}_2)^{2}, \quad \hat{k}=k/|k|
                                                \label{pol2}
\end{eqnarray}
and 
$$\sum_{\ell_{1},\ell_{2}=1,2} ( (k_{1}\times e(k_{1},\ell_{1}), 
     (k_{2}\times e(k_{2},\ell_{2}))^{2}
 = |k_{1}|^{2}|k_{2}|^{2}
     (1+(\hat{k}_{1},\hat{k}_{2})^{2})
$$
we can directly see that 
\begin{eqnarray}
&&
\hspace{-1cm}
\ce_{0}= \frac{1}{4}
    \ikm  
   \left\{
            \frac{4|k_{2}|^{2}}{\ce(k_{1})^{2}\ce(k_{1},k_{2})}
          -\frac{4|k_{2}|^{2}}{\ce(k_{1})^{2}\ce(k_{2})}
          +\frac{(k_{1},k_{2})(1+(\hat{k}_{1}, \hat{k}_{2}))}
            {\ce(k_{1})\ce(k_{2})\ce(k_{1},k_{2})} 
       \right \}  \label{E0} \\
&&
\hspace{-1cm}
\ce_{3}=\frac{1}{16}
         \ikm         \lkk 
         \frac{4|k_{1}|^4|k_2|^2}
                              {\ce(k_{1})^{4}\ce(k_{1},k_{2})}
       - \frac{4|k_{1}|^4|k_{2}|^{2}}
            {\ce(k_{1})^{4}\ce(k_2)}\right. \nonumber \\
&&\hspace{2cm} \left. + \frac{\lk -2|k_{1}|^{2}|k_{2}|^{2}
(1-(\hat{k}_{1},\hat{k}_{2})^{2})+\AB{\skk}\rk (k_{1}, k_{2})}
                              {\ce(k_{1})\ce(k_{2})\ce(k_{1},k_{2})^{3}}
              \right\},  \nonumber \\
&&
\hspace{-0.5cm}=\frac{1}{16}
         \ikm         \lkk 
         \frac{4|k_{1}|^4|k_2|^2}
                              {\ce(k_{1})^{4}\ce(k_{1},k_{2})}
       - \frac{4|k_{1}|^4|k_{2}|^{2}}
                     {\ce(k_{1})^{4}\ce(k_2)}\right. \nonumber \\
&&\hspace{4cm} \left. + \frac{-2|k_{1}|^{2}|k_{2}|^{2}
(1-(\hat{k}_{1},\hat{k}_{2})^{2})(k_{1}, k_{2})}
                              {\ce(k_{1})\ce(k_{2})\ce(k_{1},k_{2})^{3}}
               \right\},  \label{E3}  \\
&&
\hspace{-1cm}
\ce_{4}=-\frac{1}{16}
         \ikm
         \lkk 
         \frac{4|k_{1}|^4|k_2|^2}
               {\ce(k_{1})^{4}\ce(k_{1},k_{2})}  \right. \nonumber \\
&&\hspace{3cm}\left. + \frac{
\lk -2|k_{1}|^{2}|k_{2}|^{2}(1-(\hat{k}_{1},\hat{k}_{2})^{2})
           +\AB{\skk}  \rk |k_{1}|^{2}}
     {\ce(k_{1})^{3}\ce(k_{2})\ce(k_{1},k_{2})}  \right\},  
                 \label{E4}
\end{eqnarray}
where 
$\d \int d{\bf k} f(k_1,k_2)= 
\int_{ {\kappa/m} 
       \leq |k_1|,|k_2|\leq {\Lambda/ m}} 
      \frac{1}{4(2\pi)^6} f(k_1,k_2)dk_1dk_2$
and 
\begin{eqnarray*}
\ce(k) &=& \frac{1}{2}|k|^{2} +\omega(k),\\
\ce(k_{1},k_{2}) &=& \frac{1}{2}|k_{1}+k_{2}|^{2} 
                       +\omega(k_{1})+\omega(k_{2}).
\end{eqnarray*}
Moreover 
$$
\skk
 \equiv \sum_{\ell_{1},\ell_2=1,2}  \sigma \cdot 
         [(k_{1}\times e(k_{1},\ell_{1}))
                 \times (k_{2}\times e(k_{2},\ell_{2}))]
      ((k_{1}\times e(k_{1},\ell_{1}))
                 \cdot(k_{2}\times e(k_{2},\ell_{2})).
$$
It is seen that $\skk=-\s(k_2,k_1)$, 
$\skk=\s(-k_1,k_2)$ and  $\skk=\s(k_1,-k_2)$, 
from which 
it follows that 
$$\d \ikm 
\frac{(k_{1}, k_{2})\AB{\skk}}
                              {\ce(k_{1})\ce(k_{2})\ce(k_{1},k_{2})^{3}}
             =0.$$
We used this fact in the computations of $\ce_3$.

%%%%%%%%%%%%%%%%%%%%%%%%%%%%%%%%%%%%%%%%%%%%%%%%%
\subsection{Origin of the $\Lambda^{2}$ Divergence of $m_{eff}$}
%%%%%%%%%%%%%%%%%%%%%%%%%%%%%%%%%%%%%%%%%%%%%%%%%
The term $\ce_3$ has  the term  ${\displaystyle -\frac{1}{2}}E_{(2)}E_{b}$
which diverges as $(\La/m)^2$ 
as $\La\rightarrow \infty$. We intuitively see that 
this divergence is canceled in the way mentioned below.  
The first term of the integrand of $\ce _3$ behaves as 
\eq{b1}
  \frac{1}{|k_1||k_2|} 
\frac{4 |k_{1}|^{4}|k_{2}|^{2}}{\ce(k_{1})^{4}\ce(k_{1},k_{2})}
  \approx    \frac{1}{|k_1||k_2|} \frac{32}{\ce(k_{1})^{2}}
\en 
for  $|k_1|/|k_2|\ll 1$, and then 
$$
\ikmm  {\frac{1}{|k_{1}||k_{2}|} \frac{32}{\ce(k_{1})^{2}}}\approx (\La/m)^2
$$
as $\La\rightarrow \infty$. 
On the other hand, this divergence is canceled by 
${\displaystyle -\frac{1}{2}}E_{(2)}E_{b}$, 
since it  has the integral kernel 
$$
\frac{1}{|k_1||k_2|}\frac{4 |k_{1}|^{4}}{\ce(k_{1})^{4}} 
     \frac{|k_{2}|^{2}}{\ce(k_{2})}
    \approx   \frac{1}{|k_1||k_2|}\frac{32}{\ce(k_{1})^{2}} 
$$ 
for $|k_1|/|k_2|\ll 1 $. Hence we shall see that $\ce_3$ diverges
as $\sqrt{\La/m}$ but not $(\La/m)^2$. 
The divergence $(\La/m)^2$ of $\ce_{0}$ is also  canceled by 
${\displaystyle -\frac{1}{2}}E_{(2)}E_{a}$.
The integrand of $\ce_4$ has the same kernel  as \kak{b1}. 
The divergence $-(\La/m)^2$ of $\ce_4$ coming from \kak{b1}, however,  
remains without being subtracted. 
Then we obtain that $\ce_4\approx - (\La/m)^2$ as $\La\rightarrow \infty$. 
The actual calculations of $\ce_{i}$ 
are done by  this idea and straightforward.

%%%%%%%%%%%%%%%%%%%%%%%%%%%%%%%%%%%%%%%%%%%%%
\subsection{Proof of Theorem \ref{main}}
%%%%%%%%%%%%%%%%%%%%%%%%%%%%%%%%%%%%%%%%%%%%%%
\bl{e4}
There exist positive constants $c_1$ and $c_2$ such that 
\eq{ee42}
-c_1\leq \lim_{\La\rightarrow \infty}\frac{\ce_4}{(\La/m)^2}\leq -c_2.
\en
\el
\bl{eee0}
It follows that 
\eq{ee0}
\lim_{\La\rightarrow \infty}\frac{\ce_{0}}{\l 2}< \infty,
\en
and 
\eq{ee3}
\lim_{\La\rightarrow \infty}\frac{\ce_{3}}{\l 2}<\infty.
\en 
\el
{\it Proof of Theorem \ref{main}}

It is obtained that 
\eqnn 
&& a_1(\La/m)=(2/3)c_1(\La/m),\non\\
&&\label{36} a_2(\La/m)=(2/3)c_2(\La/m)+(2/3)^2c_1(\La/m)^2
\ennn 
in Theorem \ref{3}. 
A direct calculation yields that 
\eqnn 
&& \label{annn} c_1(\La/m)=\mmm((A_\mu+\pf\z\bs)\g 0, \z (A_\mu+\pf\z\bs)\g 0)_\hhh\\
&& =\mmm(A_\mu\g 0, \z A_\mu\g 0)_\hhh +
\mmm (\pf_\mu\z\bs \g 0, \z \pf_\mu\z\bs\g 0)_\hhh \non \\
&&=2 \int\frac{\vp_m(k)}{2(2\pi)^3\omega(k)}
         \frac{1}{\half|k|^2+\omega(k)}dk+
              \frac{1}{4}\int\frac{\vp_m(k)}{2(2\pi)^3\omega(k)}
              \frac{2|k|^2}{(\half|k|^2+\omega(k) )^3}dk\non \\
&&=\frac{4\pi}{(2\pi)^3} \int_{\kappa/m}^{\La/m}\frac{r}{\half r^2+r}dr
   +\frac{1}{4} \frac{4\pi}{(2\pi)^3} 
    \int_{\kappa/m}^{\La/m}\frac{r^5}{(\half r^2+r)^3} dr.\non 
\ennn 
Thus we have 
\eq{spk}
a_1(\La/m)=\frac{2}{3}
\frac{4\pi}{(2\pi)^3} 
\lk 
    \int_{\kappa/m}^{\La/m}\frac{r}{\half r^2+r} dr +
        \frac{1}{4} 
        \int_{\kappa/m}^{\La/m}\frac{r^5}{(\half r^2+r)^3} dr\rk
\en 
and there exist positive constants $b_1$ and $b_2$ such that 
\eq{bb}
b_1\leq \liml\frac{a_1(\La/m)}{\log(\La/m)}\leq b_2.
\en 
Our analysis in  Lemmas \ref{divless}, \ref{e4} and \ref{eee0} implies that 
there exist constants $c_1$ and $c_2$ such that 
$$-c_1\leq \liml\frac{2\Re \RRR{16}{1}{2}}{(\La/m)^2}\leq -c_2$$
and 
$$\liml\frac{\RRR i j k}{(\La/m)^2}=0$$
for $(i,j,k)\not=(16, 1, 2)$. Thus by \kak{76} we have 
\eq{cc}
-c_1\leq \liml\frac{c_2(\La/m)}{(\La/m)^2}\leq -c_2.
\en 
Hence 
$$-c_1\leq \liml \frac{a_2(\La/m)}{(\La/m)^2}\leq -c_2$$ 
follows 
from \kak{36}, \kak{bb} and \kak{cc}. 
\qed
From \kak{annn} we see that $e^2$ order of the effective mass, $\mass^{\rm spinless}$, for the spinless Pauli-Fierz Hamiltonian is 
$\d \frac{2}{3}\frac{4\pi}{(2\pi)^3} \int_{\kappa/m}^{\La/m}\frac{r}{\half r^2+r} dr$. 
By \kak{spk} we have a corollary. 
\bc{cor}
\label{p1}
It follows that 
$\mass>\mass^{\rm spinless}$ 
for $e$ with a sufficiently small $|e|$ but $\not=0$. 
\ec
\proof 
We have 
$$\left. \frac{d}{de^2} (\mass-\mass^{\rm spinless}) \right\lceil_{e^2=0}= 
\frac{1}{6} \frac{4\pi}{(2\pi)^3} 
\int_{\kappa/m}^{\La/m}\frac{r^5}{(\half r^2+r)^3} dr>0.$$
Since $\mass=\mass^{\rm spinless}$ for $e=0$ and both of $\mass$ and $\mass^{\rm spinless}$ are analytic in $e^2$, the corollary follows. \qed

%%%%%%%%%%%%%%%%%%%%%%%%%%%%%%%%%%%%%%%%%%%%%%%%%%%%%%%%%
\subsection{Proof of Lemmas \ref{e4} and \ref{eee0}}
%%%%%%%%%%%%%%%%%%%%%%%%%%%%%%%%%%%%%%%%%%%%%%%%%%%%%%%%%
To prove Lemmas \ref{e4} and \ref{eee0}, we prepare some notation. 
To estimate  $\ce_{i}$, $i=0,3,4$, we introduce the polar coordinate 
$(r_1,    \theta,      \phi,    r_2,      \theta_2,     \phi_2)$, where 
$r_{i}=|k_{i}|$, $i=1,2$, $0\leq \theta_2\leq \pi$, 
$0\leq \phi_2\leq 2\pi$, 
$\cos \theta=(\hat{k}_{1},\hat{k}_{2})$, $0\leq \theta\leq \pi$, 
and 
$0\leq \phi\leq 2\pi$. 
We introduce 
\eqn 
&& \ccr =\ccr(r_1,r_2)\equiv {1\over2}(r_{1}^{2}+r_{2}^{2})+r_{1}+r_{2}.\\
&& \ce_{\pm}=\ce_{\pm}(r_{1},r_{2})
    \equiv  \ccr \pm r_{1}r_{2}
     = \frac{1}{2}(r_{1}\pm r_{2})^{2}+r_{1}+r_{2}
\enn 
Then 
$\ce(k_{1},k_{2})=\ccr(r_1,r_2)+r_{1}r_{2}\cos \theta$. We set 
\eqn 
&& 
\KK=\KK(r_1,r_2)
   \equiv   \int_{0}^{\pi}   \frac{1}
    {\ccr+r_{1}r_{2}\cos \theta}r_{1}r_{2}\sin\theta d\theta,\\
&& \KKK=\KKK(r_1,r_2)
     \equiv  \int_{0}^{\pi}  \frac{(1-\cos^{2}\theta)\cos \theta}
     {(\ccr+r_{1}r_{2}\cos \theta)^3}r_{1}r_{2}\sin\theta d\theta.
\enn
Then we have 
\begin{eqnarray*}
\KK&=& \log \frac{\ce_{+}}{\ce_{-}}
                 =\log\frac{{\cal R}+r_1r_2}{{\cal R}-r_1r_2},\\
\KKK &=&-\frac{1}{r_{1}^{3}r_{2}^{3}}
          \left[6r_{1}r_{2}+
            \frac{2r_{1}^{3}r_{2}^{3}}{\ce_{+}\ce_{-}}
                 -3{\cal R}\log\frac{\ce_{+}}{\ce_{-}} \right]
\end{eqnarray*}
from which it follows that for $\zeta\equiv r_{1}r_{2}/{\cal R}<1 $, 
\begin{equation}
\KK(r_{1}, r_{2})=2\sum_{n=0}^{\infty} \frac{1}{2n+1}
               \zeta^{2n+1},\quad
\KKK(r_{1}, r_{2})
              =-\frac{1}{R^{2}}\left[\frac{4}{5} \zeta^{2}
     +{\mathcal O}\left(\zeta^{4}\right ) \right ]
\end{equation}

By the Taylor expansion of $\KK(r_1, r_2)=\KK (r_{1},r_{1}/x)$ and 
${2r_1r_2}/{\ce(r_2)}=4x(1+(2x/r_1))^{-1}$ by 
$x=r_{1}/r_{2}$ 
(note that $\zeta=2x-4(1+r_{1}^{-1})x^{2}+{\mathcal O}(x^{3})$),
we have for $r_1/r_2<1$ and $r_1\gg 1$, 
\eqnn
\label 
{ii}
\KK(r_{1},r_2) 
  &=&   4 x-\frac{8}{r_1}x^2+(\frac{4}{3}+{\mathcal O}(\frac{1}{r_1})) x^3
+{\mathcal O}(x^4), \\
\frac{2r_1r_2}{\ce(r_2)}
&=& \label{iiii}
4x-\frac{8}{r_1}x^2+\frac{16}{r_1^2}x^3+{\mathcal O}(\frac{x^4}{r_1^3}).
\ennn 
From \kak{ii} and \kak{iiii} it follows that  for $r_1/r_2<1$ and $r_1\gg 1$, 
\begin{eqnarray}
\KK(r_{1},r_{2})-\frac{2r_{1}r_{2}}{\ce(r_{2})}
&=&(\frac{4}{3}+{\mathcal O}(\frac{1}{r_1})) \frac{r_1^3}{r_2^3}+
{\mathcal  O}(\frac{r_1^4}{r_2^4}). \label{i}
\end{eqnarray}

%%%%%%%%%%%%%%%%%%%%%%%%%%%%%%%%%%%%%%%%%%%%%%%%%%%%%
\subsubsection{Proof of Lemma \ref{eee0}}
%%%%%%%%%%%%%%%%%%%%%%%%%%%%%%%%%%%%%%%%%%%%%%%%%%%
\proof 
For notational simplicity  we set $m=\kappa=1$. Replacing 
$(k_1,k_{2})$ by $|k_{1}||k_{2}|$ in (\ref{E0}), we have 
\eqnn  
&&\hspace{-1.3cm}|\ce_0 | \leq 
\frac{1}{4}
    \ikm  
   \lk 
         \frac{4|k_{2}|^{2}}{\ce(k_{1})^{2}\ce(k_{1},k_{2})}
        -\frac{4|k_{2}|^{2}}{\ce(k_{1})^{2}\ce(k_{2})}
        +\frac{2|k_1||k_2|}
        {\ce(k_{1})\ce(k_{2})\ce(k_{1},k_{2})} 
       \rk \non \\
&& \hspace{-1.3cm}= 
\label{pl} 
\frac{1}{2(2\pi)^4}
              \int_1^\La\mbox{\hspace{-2mm}}\int_1^\La dr_1dr_2 
\lkk   \frac{4r_{2}^{2}}{\ce(r_{1})^{2}}\lk  \KK(r_{1},r_{2})
   - \frac{2r_{1}r_{2}}{\ce(r_{2})}\rk
   + \frac{2r_{1}r_{2}}{\ce(r_{1})\ce(r_{2})}\KK(r_{1},r_{2})\rkk. 
\ennn 
In \cite[Lemma 4.2 (1) ]{hisp3} it is shown that for the second 
integrand of \kak{pl},  
\eq{ll1}
\int_1^\La\int_1^\La dr_1dr_2 
    \frac{2r_{1}r_{2}}{\ce(r_{1})\ce(r_{2})}\KK(r_{1},r_{2})
\leq C \log \La,
\en 
where and hereafter, $C$ denotes a non-zero constant which may be 
different from line to line. We decompose the integral region 
${\cal I}\equiv \{(r_1,r_2)| 1\leq r_1,r_2\leq \La\}$ 
into three regions:
\eqn
 {\rm  I} &=& \{(r_{1},r_{2}) \in {\cal I}| 
                  \lambda^{-1} < r_{1}/r_{2}<\lambda\},\\
 {\rm II_1}&=& \{(r_{1},r_{2})\in {\cal I}  | 
                    r_{1}/r_{2}\leq \lambda^{-1}\},\\
 {\rm II_2}&=& \{(r_{1},r_{2}) \in {\cal I} | 
                    \lambda\leq r_{1}/r_{2}\},
\enn
where $\lambda\geq 3$ is a fixed positive constant. 
By $1/\ce(r)\leq 2/r^2$ and \kak{i},we have 
$$ \left| 
\frac{4r_{2}^{2}}{\ce(r_{1})^{2}}\lk  \KK(r_{1},r_{2})
   - \frac{2r_{1}r_{2}}{\ce(r_{2})}\rk\right| 
 \leq C\frac{1}{r_1r_2}, \ \ \  (r_1,r_2)\in{\rm II_1},$$ 
and by 
$\KK(r_1,r_2)\leq C\log(r_1+r_2+1)$, 
$$
 \left| 
\frac{4r_{2}^{2}}{\ce(r_{1})^{2}}\lk  \KK(r_{1},r_{2})
   - \frac{2r_{1}r_{2}}{\ce(r_{2})}\rk\right| 
 \leq C \lk \frac{r_2^2}{r_1^4}
\log \left[ 
r_1(1+\frac{1}{r_1}+\frac{r_2}{r_1})\right] 
  +\frac{r_2}{r_1^3}\rk,\ \ \ \ (r_1,r_2) \in \rm II_2\cup I.
$$ 
Thus 
\eq{ll2}
 \int_{{\rm II_1}}  dr_1dr_2  \left| 
\frac{4r_{2}^{2}}{\ce(r_{1})^{2}}\lk  \KK(r_{1},r_{2})
   - \frac{2r_{1}r_{2}}{\ce(r_{2})}\rk\right| 
\leq C\int_1^\La dr_2 \int_1^{r_2/\lambda} dr_1\frac{1}{r_1r_2}
 \leq C[\log\La]^2, 
\en 
\eq{ll3}
\int_{{\rm I}}  dr_1dr_2  \left| 
\frac{4r_{2}^{2}}{\ce(r_{1})^{2}}\lk  \KK(r_{1},r_{2})
   - \frac{2r_{1}r_{2}}{\ce(r_{2})}\rk\right| 
\leq C\int_1^\La dr_1\int_{r_1/\lambda}^{\lambda r_1} dr_2 
(\frac{\log r_1}{r_1^2}+\frac{1}{r_1^2})  
\leq C[\log \La]^2
\en 
and 
\eq{ll33}
\int_{{\rm II_2}}  dr_1dr_2  \left| 
\frac{4r_{2}^{2}}{\ce(r_{1})^{2}}\lk  \KK(r_{1},r_{2})
   - \frac{2r_{1}r_{2}}{\ce(r_{2})}\rk\right| 
\leq C\int_1^\La dr_1\int_1 ^{r_1/\lambda } dr_2 
(\frac{\log r_1}{r_1^2}+\frac{1}{r_1^2})  
\leq C[\log \La]^2
\en 
Hence 
\kak{ee0} follows from  \kak{ll1},\kak{ll2},  \kak{ll3} and \kak{ll33}. 
In the similar way as that of $\ce_0$, we can prove \kak{ee3}. 
It is immediate to see that $0<\ce_{3}$ is bounded from above by
\eqnn 
&& \frac{1}{16}
         \ikm         \lkk 
         \frac{4|k_{1}|^4|k_2|^2}
                              {\ce(k_{1})^{4}\ce(k_{1},k_{2})}
      - \frac{4|k_{1}|^4|k_{2}|^{2}}
             {\ce(k_{1})^{4}\ce(k_2)}
      -\frac{2|k_{1}|^{3}|k_{2}|^{3}(1-(\hat{k}_{1},\hat{k}_{2})^{2})
    (\hat{k}_{1},\hat{k}_{2})}
              {\ce(k_{1})\ce(k_{2})\ce(k_{1},k_{2})^{3}}
               \right\}\non \\
&&\label{pll} 
\leq 
\frac{1}{8(2\pi)^4}\int_1^\La\int_1^\La dr_1dr_2 
\lkk    \frac{4r_{1}^{4}r_{2}^{2}}{\ce(r_{1})^{4}}
     \lk \KK(r_{1},r_{2})
   -\frac{2r_{1}r_{2}}{\ce(r_{2})} \rk
     -8 r_{1}r_{2}\KKK (r_{1},r_{2})\rkk
\ennn
where $-\KKK (r_{1},r_{2})>0$ should be noted.  For the second integrand of 
\kak{pll} it can be seen in the similar manner as 
\cite[Lemma 4.2 (4)]{hisp3} that 
\eq{ll4}
\int_1^\La\int_1^\La dr_1dr_2  r_1r_2
              \left |\KKK(r_1,r_2) \right |
        \leq C \int_1^\La\int_1^\La dr_1dr_2 \frac{1}{\ce_{-}(r_{1},r_{2})}
        \leq C \sqrt{\La}.
\en 
Note that 
$$   \left|
\frac{4r_{1}^{4}r_{2}^{2}}{\ce(r_{1})^{4}}
     \lk \KK(r_{1},r_{2})
   -\frac{2r_{1}r_{2}}{\ce(r_{2})} \rk\right|\leq 
4 \left| 
\frac{4r_{2}^{2}}{\ce(r_{1})^{2}}\lk  \KK(r_{1},r_{2})
   - \frac{2r_{1}r_{2}}{\ce(r_{2})}\rk\right|.$$
Then 
\eq{ll5}
\int_1^\La\int_1^\La 
\left|
\frac{4r_{1}^{4}r_{2}^{2}}{\ce(r_{1})^{4}}
     \lk \KK(r_{1},r_{2})
   -\frac{2r_{1}r_{2}}{\ce(r_{2})} \rk\right|\leq C[\log\La]^2
\en 
follows from \kak{ll2},  \kak{ll3} and \kak{ll33}.  
Hence \kak{ee3} follows from \kak{ll4} and \kak{ll5}. 
\qed
%%%%%%%%%%%%%%%%%%%%%%%%%%%%%%%%%%%%%%%%%%%%%
\subsubsection{Proof of Lemma \ref{e4}}
%%%%%%%%%%%%%%%%%%%%%%%%%%%%%%%%%%%%%%%%%%%%%%
\proof 
Let us decompose $\ce_4$ as 
$\ce_4=\ce_{41}+\ce_{42}$, where 
\eqn 
&&\ce_{41}=-\frac{1}{16}
\ikm \frac{4|k_{1}|^4|k_2|^2}{\ce(k_{1})^{4}\ce(k_{1},k_{2})},\\
&&\ce_{42}=
-\frac{1}{16}
         \ikm
\frac{
\lk -2|k_{1}|^{2}|k_{2}|^{2}(1-(\hat{k}_{1},\hat{k}_{2})^{2})
           +\AB{\skk}  \rk |k_{1}|^{2}}
{\ce(k_{1})^{3}\ce(k_{2})\ce(k_{1},k_{2})}. 
\enn 
We have the inequality 
$$|\ce_{42}|\leq \frac{1}{16}
         \ikm
         \lk\frac{ 6|k_{1}|^{4}|k_{2}|^{2}}
{\ce(k_{1})^{3}\ce(k_{2})\ce(k_{1},k_{2})}  \rk=
\frac{1}{16}\frac{(4\pi)(2\pi)}{(2\pi)^6}\int_1^\La\int_1^\La dr_1dr_2 
\frac{\KK (r_{1},r_{2})}{\ce(r_1)}.
$$
Here we used the inequality 
$|\AB{\skk}|\leq 4|k_1|^2|k_2|^2$. 
It is seen that 
\eq{ll6}
\int_1^\La\int_1^\La dr_1dr_2 \frac{\KK (r_{1},r_{2})}{\ce(r_1)}\leq C[\log\La]^2.\en 
We shall estimate $\ce_{41}$.  
We have 
$$\ce_{41}=
\frac{1}{16}\frac{(4\pi)(2\pi)}{(2\pi)^6}\int_1^\La\int_1^\La dr_1dr_2 
\frac{4r_1^2r_2^2}{\ce(r_1)^4}\KK(r_1,r_2).$$
Since
$$\frac{4r_1^2r_2^2}{\ce(r_1)^4}\KK(r_1,r_2)\leq C\frac{r_2^2}{r_1^6}\log(r_1+r_2+1),$$
we have 
\eq{dd1}
\int_{{\rm I}\cup {\rm II_2}}dr_1dr_2 \frac{4r_1^2r_2^2}{\ce(r_1)^4}\KK(r_1,r_2)<\infty.
\en
Moreover 
$$\KK(r_1,r_2)\leq C\frac{r_1}{r_2},\ \ \ (r_1,r_2)\in {\rm II_1},$$
we have 
\eq{dd2}
\int_{{\rm II_1}} dr_1dr_2 \frac{4r_1^2r_2^2}{\ce(r_1)^4}\KK(r_1,r_2)<
C\int_1^\La dr_2 \int_1^{r_2/\lambda} dr_1\frac{r_2}{r_1^5}\leq 
C\La^2.
\en 
Hence by \kak{ll6}, \kak{dd1} and \kak{dd2},   
\eq{ll7} 
\ce_{4} \leq C\La^2
\en 
follows. 
In the region ${\rm II_1}$ it follows that 
\eq{ll8}
4\frac{r_1}{r_2}- C\lk \frac{r_1}{r_2}\rk^2  \leq \KK(r_1,r_2).
\en 
Since 
${2}/{3}\leq {r^2}/{\ce(r)}\leq 2$, 
we see  together with \kak{ll8} that 
$$
\frac{4r_1^2r_2^2}{\ce(r_1)^4}\KK(r_1,r_2)\geq 
C\lkk \frac{r_1}{r_2}- \lk \frac{r_1}{r_2}\rk^2 \rkk 
\frac{r_2^2}{r_1^6}=C\frac{r_2}{r_1^5}-C\frac{1}{r_1^4},\ \ \ (r_1,r_2)\in{\rm II_1}.$$
Then we have 
$$ 
\int_{\rm II_1} dr_1 dr_2 \frac{4r_1^2r_2^2}{\ce(r_1)^4}\KK(r_1,r_2)
\geq C  \int_{1}^{\Lambda}dr_{1} \int_{\lambda r_{1}}^{\Lambda}dr_{2}
\lk \frac{r_2}{r_1^5}-\frac{1}{r_1^4}\rk 
 \geq C (\La^2-\La),
$$
which implies that 
\eq{ll9}
 C\La^2\leq\ce_4.
\en 
Hence the lemma follows from \kak{ll7} and \kak{ll9}. 
\qed

%%%%%%%%%%%%%%%%%%%%%%%%%%%%%%%%%%%%%%%%%%
\section{Conclusion and Discussion}
%%%%%%%%%%%%%%%%%%%%%%%%%%%%%%%%%%%%%%%%%

We have shown that the second order $a_{2}(\Lambda/m)$ of 
$\mass/m=\sum_{n=0}^\infty a_n(\La/m) e^{2n}$ diverges like $-(\La/m)^{2}$, 
which seems not to be renormalized by the conventional renormalization 
scheme of the field theory. The traditional power counting theorem means that 
this quantity diverges like $\l 2$ at most. However, in the 
present case, the integrands are not symmetric in  momenta
$k_{i}$, contrary to the case of the covariant perturbation theory.
This seems to be a reason why the conventional power counting 
theorem does not hold.

As was discussed in \cite{arai1}, the Pauli-Fierz model is obtained from 
the Maxwell-Dirac Hamiltonian 
\begin{equation}
H_{\rm DM}=\mmm\alpha_{\mu}(p-eA)_{\mu}+\beta  m +V +\hf
        \label{DM}, 
\end{equation}
where $V$ is a suitable scalar potential, and $(\alpha_1,\alpha_2,\alpha_3,\beta)$ 
are $4\times 4$ symmetric matrices satisfying anti-commutation relations. 
This Hamiltonian is self-adjoint if suitable cutoffs are introduced.
To get the scaling limit, we let $\kappa>0$ be the speed of light, 
and define
\begin{equation}
H_{\rm DM}^{(\kappa)}=\kappa \mmm\alpha_{\mu}(p-eA^{(\kappa)})_{\mu}
     +\kappa^{2}\beta  m -\kappa^{2}m +V_{0,\kappa} +  \hf
        \label{LimitDM}, 
\end{equation}
where $-\kappa^{2}m$ is introduced to ensure the ground state 
energy is zero except for negative energies of Dirac's sea of positrons. 
Moreover suitable cutoffs depending on $\kappa$ are introduced into 
$A_{\mu}$ and $V_{0}$ so that $A_{\mu}^{(\kappa)}$ turns out to be
 bounded as $\|A_{\mu}^{(\kappa)}\|<K(\kappa)$, 
which  is achieved by truncating the spectrum of  $|A_{\mu}|$ and 
$|V_{0}|$ more than $K(\kappa)$. 
Then it is proven that the scaling limit of $H_{\rm DM}^{(\kappa)}$ 
converges to the Pauli-Fierz Hamiltonian  with spin $1/2$ 
in the strong resolvent sense. So it is not clear to what extent 
the original properties of the model are kept in this limit.
Though it is not yet proven and this Hamiltonian may not 
have a  ground state as the quantum field model of $\phi_{4}^{3}$, 
we expect that $H_{\rm DM}$ is renormalizable 
since the Feynman diagrams appearing in this model are a subset of 
QED in which no fermion-loop diagrams appear.
These arguments will lead us to ask a number of questions: 
\begin{description}
\item{(1)} Is the so-called covariant renormalization method in QED 
    directly related to the renormalization of the Hamiltonian? 
\item{(2)} Renormalizability of QED seems to be ensured by an indefinite 
    metric. Does the introduction of the indefinite metric \cite{ito} 
    cures the renormalizability of the Pauli-Fierz model ?
\item{(3)} Is  the model (\ref{DM}) renormalizable ?
\end{description}

{\it Acknowledgements.} 
{\footnotesize We thank H. Spohn and E. Seiler
for helpful comments.
K. R. I. thanks Grant-in-Aid  for Science Research (C) 15540222 from MEXT and 
F. H.  thanks Grant-in-Aid  for Science Research (C) 15540191 
from MEXT. A part of the calculation was done while K.R.I. was staying 
in Max-Planck-Inst. Physik (2004, August). K.R.I. thanks Prof. E. Seiler 
for the kind hospitality extended to  him. 
}

\appendix

\newtheorem{thmA}{Theorem A}
\newtheorem{lemA}[thmA]{Lemma A}
\newtheorem{remA}{Remark A}

%%%%%%%%%%%%%%%%%%%%%%%%%%%%%%%%%%%%%%%%%%%%%%%%%%%
\section {Sketch  of the Calculations of Lemma \ref{divless}}
%%%%%%%%%%%%%%%%%%%%%%%%%%%%%%%%%%%%%%%%%%%%%%%%%%%
To estimate ${\cal E}=(\Phi_{(\ell)},H_{(m)}\Phi_{(n)})$, 
we represent $A_{\mu}$ and $\sigma B_{\mu}$ in terms of 
$a^{\#}(k,\ell)e_{\mu}(k,\ell)$ and 
$a^{\#}(k,\ell)[\sigma\cdot (i k\times e(k,\ell))]$,  and 
use the pull-through formulae 
\begin{eqnarray*}
a(k,\ell) ({\half }\pf^{2}+\hf)^{-1}
    &=&({\half}(\pf+k)^{2}+\omega(k)+\hf)^{-1}a(k,\ell),\\
a(k,\ell)\pf_\mu &=&(\pf_\mu +k_{\mu})a(k,\ell)
\end{eqnarray*}
and canonical commutation relations: 
$$[a(k_{1},\ell_{1}), a^\ast  (k_{2},\ell_{2})]
=\delta(k_{1}-k_{2})\delta_{\ell_{1},\ell_{2}}.
$$
The terms containing no $\s B$ have been estimated in \cite{hisp3}. 
The terms containing two $\sigma B$ consist of 
two $A_{\mu}$'s, two $\sigma B$'s  and 0 or two or four $\pf$'s.
Let $k_{i}, e(k_{i},\ell_{i})$, $i=1,2$,   be two independent 
momenta and polarizations vectors of photons.  Then using 
(\ref{spin1})--(\ref{pol2}), we have the following factors $P(k_{i})$ 
coming from the polarization vectors $e_{i}=e(k_{i},\ell_{i})$, 
the Pauli-spin matrices $\sigma\cdot(k_{i}\times e_{i})$, and 
$k_{i}$ coming from $\pf$: 
\begin{eqnarray*}
&& \sum_{\ell_1,\ell_2=1,2} [\sigma\cdot (k_{1}\times e_{1})]^{2}(e_{2},e_{2}) 
          =4|k_{1}|^{2} \\
&& \sum_{\ell_1,\ell_2=1,2} [\sigma\cdot (k_{1}\times e_{1})] [\sigma(k_{2}\times e_{2})]
     (e_{1},e_{2})  =2(k_{1},k_{2}) +i\sigma\cdot (\cdots)\\
&& \sum_{\ell_1,\ell_2=1,2} [\sigma\cdot (k_{1}\times e_{1})] [\sigma(k_{2}\times e_{2})]
     (k_{2},e_{1})(k_{1},e_{2}) 
   =-(k_{1},k_{2})(1-(\hat{k}_{1},\hat{k}_{2})^{2}) +i\sigma\cdot (\cdots)
\end{eqnarray*}
The sum of the contributions from $i\sigma(\cdots)$ is zero.
We have 
${\cal E}^{p}(k_{1},k_{2}){\cal E}^{q}(k_{1}){\cal E}^{r}(k_{2})$ 
in the denominator coming from $H_{(1)}=H_{0}^{-1}$ where
$p+q+r\geq 3$ is equal to the number of $H_{0}^{-1}$. Let $q'$ and $r'$ 
be the numbers  of $\sigma B$ and $\pf$. Then 
$p+q+r-(q'+r')/2=3$ and $q'=0,2,4$,  and we see that ${\cal E}$ 
is represented as  sum of the following integrals: 
\begin{equation}
\label {ll}
\frac{1}{4(2\pi)^{6}}
  \int\frac{1}{\omega(k_{1})\omega(k_{2})}
\frac{P(k_{i})}
{{\cal E}^{p}(k_{1},k_{2}) {\cal E}^{q}(k_{1}){\cal E}^{r}(k_{2})}dk_{1}dk_{2}
\end{equation}
where $P(k_i)$ is a polynomial of $k_{i}$ of degree $q'+r'$.
We estimate \kak{ll}. Changing variables $(k_1, k_2)$ in \kak{ll}
to the polar coordinate stated in the proofs of Lemmas \ref{e4} and \ref{eee0}, 
and integrating with respect to $0\leq \phi, \phi_2\leq 2\pi$ and $0\leq\theta_2\leq \pi$,  
we have $dk_{1}=4\pi r_{1}^{2}dr_{1}$ and 
$dk_{2}=2\pi r_{2}^{2} \sin \theta dr_{2} d\theta$ so that 
$(k_{1},k_{2})=r_{1}r_{2}\cos \theta$, $r_{i}=|k_{i}|$. Then we see 
that the integral over $0<\theta<\pi$ takes one of the following forms:
\begin{eqnarray*}
{\cal K}_{1,p} &=& \int_0^\pi \frac{1}{({\cal R}+r_{1}r_{2}\cos\theta)^{p}}
                      r_{1}r_{2}\sin \theta d\theta, \\
\hat{\cal K}_{1,p} &=& \int_0^\pi
                   \frac{\cos}{({\cal R}+r_{1}r_{2}\cos\theta)^{p}}
                   r_{1}r_{2}\sin \theta d\theta,\\
{\cal K}_{2,p} &=& \int_0^\pi 
                  \frac{1-\cos^{2}\theta}{({\cal R}+r_{1}r_{2}\cos\theta)^{p}}
                   r_{1}r_{2}\sin \theta d\theta, \\
\hat{\cal K}_{2,p} &=& \int_0^\pi \frac{(1-\cos^{2}\theta)\cos\theta}
              {({\cal R}+r_{1}r_{2}\cos\theta)^{p}}
                         r_{1}r_{2}\sin \theta d\theta
\end{eqnarray*}
where we recall ${\cal R}={1\over2}(r_{1}^{2}+r_{2}^{2})+r_{1}+r_{2}$. 
Then  we have 
\begin{eqnarray*}
{\cal  K}_{1,p}(r_{1},r_{2}) &=&
\left \{ \begin{array}{lr}
    \displaystyle   \log \frac{{\cal E}_{+}}{{\cal E}_{-}}  &  p=1 \\
    \displaystyle \frac{1}{p-1}
    \left(\frac{1}{{\cal E}_{-}^{p-1}}-
         \frac{1}{{\cal E}_{+}^{p-1}}\right)  &  p>1 
         \end{array}
 \right.  \\[2ex]
{\cal K}_{2,p}(r_{1},r_{2})
&=& \frac{1}{(r_{1}r_{2})^{2}}
 \times 
\left \{
  \begin{array}{lr} 
\displaystyle
        \left[ -{\cal E}_{+} {\cal E}_{-} \log\frac{\cal E_{+}}{\cal E_{-}}
        +2{\cal R}r_{1}r_{2}  \right]     & p=1 \\[3ex]
\displaystyle
         \left [ 2{\cal R} \log\frac{\cal E_{+}}{ \cal E_{-}}
        -4r_{1}r_{2}  \right ]     & p=2 \\[3ex]
\displaystyle
         \left [ \frac{2{\cal R} r_{1}r_{2}}{{\cal E}_{+}{\cal E}_{-}}
          - \log\frac{\cal E_{+}}{ \cal E_{-}}
          \right ]     & p=3 \\
   \end{array}
\right . 
\end{eqnarray*}
\begin{eqnarray*}
\hat{\cal K}_{1,p}(r_{1},r_{2}) &=&
        \frac{1}{r_{1}r_{2}} \times 
       \left \{ \begin{array}{lr}
       \displaystyle  \left[2r_{1}r_{2}
                    -{\cal R} \log\frac{\cal E_{+}}{\cal E_{-}}\right] 
                      & p=1  \\[3ex]
       \displaystyle  \left[ \log \frac{\cal E_{+}}{\cal E_{-}} 
       -{\cal R} \left(\frac{1}{\cal E_{-}}-\frac{1}{\cal E_{+}}\right)\right]
                       & p=2 
                \end{array}
        \right . \\
\hat{\cal K}_{2,3}(r_{1},r_{2}) &=&
    \displaystyle \frac{1}{(r_{1}r_{2})^{3}}
       \left [3{\cal R}\log \frac{\cal E_{+}}{\cal E_{-}} -6r_{1}r_{2}
       -  \frac{2r_{1}^{3}r_{2}^{3}}{{\cal E}_{+}{\cal E}_{-}}\right ]
\end{eqnarray*}
where ${\cal E}_{\pm}={\cal R}\pm r_{1}r_{2}$. 
They have the following Taylor expansions by 
$\zeta=r_{1}r_{2}/{\cal R}$:
\begin{eqnarray*}
{\cal K}_{1,p}(r_{1},r_{2}) &=&
\left \{ \begin{array}{lr}
     \displaystyle  \sum_{n=0}^{\infty} \frac{2}{2n+1}
                     \zeta^{2n+1}  &  p=1 \\
    \displaystyle \frac{-2}{p-1} \,
    \frac{1}{{\cal R}^{p-1}}
     \sum_{n=0}^{\infty}\left(
                              \begin{array}{c}
                              -p+1 \\
                              2n+1 
                             \end{array}
                           \right)
                             \zeta^{2n+1}
                               & p>1 
          \end{array}
         \right .       \\[2ex]
{\cal K}_{2,p}(r_{1},r_{2})
&=&  
\left \{
  \begin{array}{lr} 
\displaystyle
   {4\over3}\zeta+{2\over3}\zeta^3
    +\sum_{n=2}^{\infty}(1-\zeta^{-2}) \frac{2}{2n+1}\zeta^{2n+1} & p=1 \\[3ex]
\displaystyle
        \frac{4}{\cal R} \sum_{n=1}^{\infty}
                \frac{1}{2n+1}\zeta^{2n-1}     & p=2 \\[3ex]
\displaystyle
          \frac{2}{{\cal R}^{2}}\sum_{n=1}^{\infty}
            \left(1-\frac{1}{2n+1}\right)\zeta^{2n-1}  & p=3
   \end{array}
\right .   \\
\end{eqnarray*}
\begin{eqnarray*}
\hat{\cal K}_{1,p}(r_{1},r_{2}) &=&
       \left \{ \begin{array}{lr}
       \displaystyle -\sum_{n=1}^{\infty}\frac{2}{2n+1}\zeta^{2n}
                      & p=1  \\[3ex]
       \displaystyle -\frac{2}{\cal R} 
                      \sum_{n=1}^{\infty}
                      (1-\frac{1}{2n+1})\zeta^{2n}
                       & p=2 
                \end{array}
        \right . \\
\hat{\cal K}_{2,3}(r_{1},r_{2}) &=&
    \displaystyle -\frac{2}{{\cal R}^{2}}
                   \sum_{n=2}^{\infty}
                  \left(1-\frac{3}{2n+1}\right)\zeta^{2(n-1)}
\end{eqnarray*}
Note that
$$\zeta=\frac{2x}{1+(2+\frac{1}{r_{1}})x+(1+\frac{1}{r_{1}})x^{2}}
       =2x -(4+\frac{2}{r_{1}})x^{2}+O(x^{3})
$$
$x=r_{1}/r_{2}$. 
Since ${\cal K}>0$ (resp. $\hat{\cal K}<0$)  are odd (resp. even)
in $\zeta$, the first terms are enough. We decompose 
$[1,\Lambda/m]\times [1,\La/m]$
into the symmetric region ${\rm I}$ and the asymmetric regions 
${\rm II}_1$ and ${\rm II}_2$ 
and we use the Taylor expansion in the regions ${\rm II}_1$ and 
${\rm II}_2$.  Then the following integrals are fundamental and 
left to the reader as an exercise:
\eqn 
&& \int _{[1,\Lambda/m]\times[1,\La/m]} \frac{1}{{\cal R}-r_{1}r_{2}}dr_{1}dr_{2}
   \approx \sqrt{\Lambda/m}, \\
&& 
\int _{[1,\Lambda/m]\times[1,\La/m]} \frac{1}{r_{1}r_{2}}
                   {\cal K}_{1,1}(r_{1},r_{2})dr_{1}dr_{2}
     \approx  \log (\Lambda/m)
\enn 
as $\La\rightarrow\infty$. 
Thus the remaining integrals with two $\s B$ are done following 
the methods described in Section 3.5. The terms with four $\s B$'s, 
i.e., $\RRR 6 3 2$ and $\RRR 6 1 6$  are similarly estimated. 

\end{document}